\documentclass[a4paper,12pt]{article}
\pdfoutput=1 

\usepackage{jheppub} 

\usepackage[T1]{fontenc} 

\usepackage{xcolor}
\usepackage{todonotes}
\usepackage{soul}

\usepackage{float}
\usepackage{subfigure}
\usepackage{multirow}

\graphicspath{{Figures/}}

\title{\boldmath Lyapunov Exponents and Phase Structure of Lifshitz and  Hyperscaling Violating Black Holes}

\author[a,b,1]{A. Naveena Kumara,\note{Corresponding author.}}
\author[c]{Shreyas Punacha}
\author[d]{and Md Sabir Ali}

\affiliation[a]{Department of Physics, Ben-Gurion University of the Negev,\\ Beer-Sheva 84105, Israel}
\affiliation[b]{Division of Theoretical Physics,  Rudjer Bo\v skovi\'c Institute,\\ Bijeni\v cka c.54,  HR-10002 Zagreb, Croatia}
\affiliation[c]{Department of Oral Health Sciences, School of Dentistry,\\ University of Washington, Seattle, WA 98195, USA.}
\affiliation[d]{Lanzhou Center for Theoretical Physics \& Key Laboratory of Theoretical Physics of Gansu Province, Lanzhou University, Lanzhou, Gansu 730000, China\\
Key Laboratory of Quantum Theory and Applications of MoE,
Lanzhou University, Lanzhou, Gansu 730000, China\\
Institute of Theoretical Physics \& Research Center of Gravitation, Lanzhou University, Lanzhou 730000, China}
\emailAdd{naviphysics@gmail.com, nathith@irb.hr}
\emailAdd{shreyasp444@gmail.com, shreyas4@uw.edu}
\emailAdd{alimd.sabir3@gmail.com}

\abstract{We study the phase structure of Lifshitz and hyperscaling violating black holes using Lyapunov exponents. For describing hyperscaling violating system, we chose a particular gravity model constructed from generalized Einstein-Maxwell-Dilaton action which includes the Lifshitz cases in appropriate limit.  We study the relationship between Lyapunov exponents and black hole phase transitions considering both the timelike and null geodesics. We observe that, the black hole phase transiton properties are reflected in Lyapunov exponent where its multiple branches correspond to the distinct phases of the black hole. The discontinuos change of the Lyapunov exponent  during the phase transition  serve as an order parameter with critical exponent $1/2$ near the critical point. Our numerical study reveals that the correlation between the Lyapunov exponent and black hole thermodynamic properties can be generalised beyond the AdS spacetime. We find that it is independent of the HSV parameter as well as the Lifshitz exponent. }


\begin{document} 
\maketitle
\flushbottom

\section{Introduction}

Gauge/gravity duality, also termed as holographic correspondence, establishes a profound relationship between a (relativistic) field theory and a gravity theory in one higher spatial dimension \cite{Susskind:1994vu, Maldacena:1997re, Gubser:1998bc, Witten:1998qj, Aharony:1999ti}. The application of this duality to real-world field theories has sparked renewed interest in the phenomenology of gravitational models. An example of this duality is the anti-de Sitter/conformal field theory (AdS/CFT) correspondence, which links string theory in negatively curved backgrounds to conformal field theories in one less non-compact spatial dimension. Notably, the strong coupling limit of the boundary field theory aligns with the supergravity limit of the bulk string theory. Consequently, holography serves as a prototype for illustrating features of strongly coupled relativistic field theories. This modeling approach has found diverse applications, including investigations into the physics of the quark/gluon plasma and exploring condensed matter systems, ranging from superconductors to (non) Fermi liquids. Despite these applications, there is a widespread effort to understand and employ holography in a more generic manner. In condensed matter physics, numerous systems are believed to be governed by strongly interacting non-relativistic physics \footnote{For instance, in the realm of field theories featuring finite charge density, the presence of an electric current disrupts Lorentz invariance.}, prompting the natural question of whether holography can provide insights into such systems (See Refs \cite{Hartnoll:2009sz, Herzog:2009xv, McGreevy:2009xe, Sachdev:2011wg} for the application of holographic techniques to the study of condensed matter systems).

Gauge/gravity duality is a useful tool for studying strongly-coupled systems near critical points. At such points, the system exhibits a scaling symmetry and can be described by a conformal field theory (CFT). From the gauge/gravity perspective, this means that the gravitational theory is defined on a metric that is asymptotically locally Anti-de Sitter (AdS). However, in many physical systems, critical points are characterized by dynamical scalings, where, despite exhibiting scaling symmetry, space and time scale differently. A notable example is the Lifshitz fixed point, where the system is spatially isotropic and scale-invariant, but time exhibits an anisotropic scaling characterized by a dynamical exponent, $z$. This scale symmetry is expressed as follows: 
\begin{equation} \label{scale}
t\rightarrow \xi^z t, \qquad x_i\rightarrow \xi x_i,    
\end{equation}
here $t$ represents time, and $x_i$ denotes spatial coordinates.  In the context of gauge/gravity duality, it is essential to find gravitational theories that provide a gravity description of Lifshitz fixed points. The corresponding metric can be written as 
\begin{equation}
    ds^2=-r^{2z}dt^2+r^2 \sum_{i=1}^d dx_i^2+\frac{dr^2}{r^2}.
\end{equation}

It is evident that the Lifshitz geometry, characterized by its anisotropic nature, cannot be a solution of pure Einstein gravity. (Notably, when  $z=1$, the metric reduces to the familiar AdS form). A gravity dual for the Lifshitz fixed point can be obtained by considering a gravitational theory that admits a solution showcasing the scaling symmetry mentioned in Eq. \ref{scale} (see Ref. \cite{Kachru:2008yh}). In the context of \textit{bottom-up} model building, a natural framework to explore is the Einstein-Maxwell-Dilaton (EMD) theory. This setup has been extensively studied in the literature, yielding Lifshitz-like black brane geometries \cite{Goldstein:2009cv, Cadoni:2009xm, Perlmutter:2010qu, Goldstein:2010aw, Bertoldi:2010ca, Bertoldi:2011zr, Cadoni:2011kv, Iizuka:2011hg, Berglund:2011cp, Tarrio:2011de, Myung:2012cb, Pal:2012zn, EslamPanah:2022ewo, Hendi:2013zba, Bravo-Gaete:2021kgt} (see Ref. \cite{Taylor:2015glc} for a review on the holographic modeling of field theories with Lifshitz symmetry). With the inclusion of a dilaton and Abelian gauge fields, more sophisticated metrics generalizing Lifshitz can be derived. These metrics, in addition to anisotropic scaling, may feature an overall hyperscaling factor. Specifically, a geometry can take the form:
\begin{equation}
    \label{metric1}
ds^2=r^{\frac{-2\theta}{d}} \left(-r^{2z}dt^2+r^2\sum_{i=1}^ddx_i^2
+\frac{dr^2}{r^2} \right),
\end{equation}
where $z$ and $\theta$ are dynamical and hyperscaling violation (HSV) \footnote{We use the abbreviation HSV for hyperscaling-violating and hyperscaling violation interchangeably} exponents, respectively. To reiterate, $z$ characterizes the departure from Lorentz invariance, while $\theta$ characterizes the deviation from the scale-invariant limit. This generalized geometry has garnered recent attention \footnote{The Lifshitz geometry has been proposed as a potential framework for understanding the behavior of strange metals in a holographic context\cite{Hartnoll:2009ns}. Additionally, to study a system that has a Fermi surface, we can examine hyperscaling violating (HSV) geometries within a specific range of parameters for its gravitational dual \cite{Dong:2012se}.}, and HSV Lifshitz solutions have been identified in various settings \cite{Singh:2010zs, Narayan:2012hk, Singh:2012un, Dey:2012tg, Dey:2012rs, Dey:2012fi,  Kastor:2018cqc}. It represents the most general spatially homogeneous and covariant geometry under the scale transformations: \footnote{When examining the gravitational perspective, the issue can be approached in reverse. As indicated in equation \ref{HSV_trans}, a distinctive characteristic of the HSV metric is that the proper distance in the emergent spacetime undergoes non-trivial transformations under scale transformations, governed by the exponent $\theta$. In the standard AdS/CFT correspondence, the proper distance remains invariant with $\theta = 0$, rather than exhibiting covariant transformation under scale operations.

A meaningful connection exists between volume elements in the holographic space and various entropic measures of the boundary theory. This connection implies that a non-zero value of $\theta$ will alter the scale transformation of the thermal entropy density, denoted as $S$. Consequently, using this rationale, it becomes evident that $\theta$ serves as the HSV exponent for the boundary theory. Thus, the non-invariance of the proper distance in the holographic theory implies violations of hyperscaling on the boundary. \cite{Huijse:2011ef}}

\begin{equation} \label{HSV_trans}
    t\rightarrow \xi ^z t, \quad r\rightarrow \xi ^{-1} r, \quad x_i\rightarrow \xi x_i, \quad  
ds_{d+2} \rightarrow \xi ^{\frac{\theta}{d}} ds_{d+2}.
\end{equation}
Notably, specific solutions for black holes that violate hyperscaling have been identified in gravitational theories incorporating higher-order gravitational corrections or additional matter fields, such as massive vector fields or a Maxwell field coupled to a dilation \cite{Charmousis:2010zz, Dong:2012se, Gath:2012pg, Alishahiha:2012qu, Gouteraux:2012yr, OKeeffe:2013xdv, Ghodrati:2014spa, Cremonini:2014gia, Ganjali:2015cba, Roychowdhury:2015fxf, Li:2016rcv, Ge:2016lyn, Cremonini:2016avj, Cremonini:2018jrx, Salvio:2013jia}.

On an alternate front, nearly fifty years  following  Hawking's discovery of black hole radiation, the thermodynamics of black holes continues to be a guiding light in the exploration of quantum gravity (see Ref \cite{Hawking:1971tu, Bekenstein:1972tm, Bekenstein:1973ur, Hawking:1974rv, Hawking:1975vcx, Hawking:1982dh} for pioneering works on black hole thermodynamics). Analogous to conventional thermal systems, black holes exhibit fundamental thermodynamic attributes such as temperature, entropy, and other thermodynamic properties. The variations between equilibrium configurations find systematic elucidation through the first law of thermodynamics. Notably, this similarity with conventional thermodynamics extends into the domain of phase transitions. With the introduction of the AdS/CFT correspondence \cite{Maldacena:1997re, Gubser:1998bc, Witten:1998qj}, there has been extensive investigation into the thermodynamics and critical behaviour of diverse AdS black holes \cite{Witten:1998zw, Cvetic:1999ne, Chamblin:1999tk, Chamblin:1999hg, Caldarelli:1999xj, Cai:2001dz, Cvetic:2001bk}. In particular, charged AdS black holes exhibit a van der Waals (vdW)-like phase transition. This transition encompasses a first-order phase transition terminating at a second-order critical point in a canonical ensemble \cite{Chamblin:1999tk, Chamblin:1999hg}, and a Hawking-Page-like phase transition in a grand canonical ensemble \cite{Cai:2001dz}. In the extended phase space, wherein the cosmological constant is regarded as thermodynamic pressure \cite{Kastor:2009wy, Dolan:2011xt, Kubiznak:2012wp}, investigations into the thermodynamics and critical phenomena of AdS black holes has revealed a spectrum of novel phenomena \cite{Wei:2012ui, Gunasekaran:2012dq, Cai:2013qga, Altamirano:2013ane, Altamirano:2013uqa, Xu:2014kwa, Frassino:2014pha, Dehghani:2014caa, Wei:2014hba, Dolan:2014vba, Hennigar:2015esa, Caceres:2015vsa, Wei:2015ana, Chakraborty:2015hna, Hendi:2016yof, Hennigar:2016xwd, Momeni:2016qfv, Hendi:2017fxp, Wang:2018xdz, Wei:2020poh}. The comprehensive understanding of black hole thermodynamics is still elusive, prompting a strong push to explore the phase structure of black holes from diverse perspectives. One recent endeavor in this direction involves the utilization of Lyapunov exponent\cite{Guo:2022kio}.

The Lyapunov exponent proves instrumental in studying chaotic dynamics within the framework of general relativity—a nonlinear dynamical theory. It serves as effective tool to analyze the spacetime perturbations and particle orbits in the vicinity of black holes. The Lyapunov exponent $\lambda$ can be used as an indicator of the separation rate between neighboring trajectories. It reflects the sensitivity of the system to the initial condition. When $\lambda>0$, it indicates a chaotic system, which means that even a slight difference in the initial conditions will lead to an exponential separation of trajectories. When $\lambda=0$ the system is stable, neighboring trajectories will maintain a distance without diverging or converging. If $\lambda<0$, the particle orbit will be asymptotically stable, resulting in the nearby trajectories tending to overlap. Extensive research has been devoted to exploring the chaotic motion of particles within various black hole spacetimes \cite{Sota:1995ms, Sota:1996cv, Kan:2021blg, Gwak:2022xje, Hanan:2006uf, Gair:2007kr, AlZahrani:2013sqs, Polcar:2019kwu, Wang:2016wcj, Chen:2016tmr, Wang:2018eui, Lu:2018mpr, Guo:2020xnf}. Specifically, investigations into particle motion near black hole horizons have revealed that the Lyapunov exponent adheres to a universal upper bound proposed within the gauge/gravity duality framework \footnote{ Maldacena, Shenker, and Stanford conjectured a universal upper limit on Lyapunov exponent $\lambda$ in the context of chaos in thermal quantum systems with a large number of degrees of freedom: $\lambda \leq \frac{2 \pi T}{\hbar}$, where $T$ represents temperature of the system \cite{Maldacena:2015waa}. Considering the quantum nature of black holes, characterized by a temperature derived from the Hawking expression $T = \frac{\hbar \kappa}{2 \pi}$, the conjectured upper bound can be refined to $\lambda \leq \kappa$, where $\kappa$ is surface gravity.} \cite{Hashimoto:2016dfz, Dalui:2018qqv}. However, counterexamples challenging this upper bound have been reported \cite{Zhao:2018wkl, Guo:2020pgq}.

The interrelation between the Lyapunov exponent and the phase structure of black holes becomes evident through their mutual connection with black hole quasinormal modes (QNMs). The Lyapunov exponents of unstable null geodesics have been found to be closely linked to the imaginary part of a specific class of quasinormal modes in black hole spacetime \cite{Cardoso:2008bp, Guo:2021enm}. Remarkably, black hole QNMs exhibit pronounced changes near phase transition points \cite{Liu:2014gvf, Momennia:2018hsm}. This observation suggests that the characteristic phenomenon of QNMs near phase transition points can be reflected by Lyapunov exponents. Therefore, Lyapunov exponent provide insights into the divergence and convergence rates of particle orbits around the equatorial plane of the black hole in the context of black hole phase transitions. From these observations, a direct relation between Lyapunov exponents and black hole phase transition was established in Ref. \cite{Guo:2022kio}. The occurrence of a phase transition can be identified by a discontinuous jump in the value of the Lyapunov exponent $\lambda$, and its difference $\Delta \lambda$ can be characterized as an order parameter. At the phase transition point, the critical exponent of $\lambda$ has been calculated as $1/2$, aligning with the circular orbit radius. This discovery has opened up a new pathway for exploring the black hole phase structure using the Lyapunov exponent \cite{Yang:2023hci, Lyu:2023sih}.

It is natural to extend the study from AdS spacetime to a more general setting, namely Lifshitz and HSV black holes. This article aims to investigate the thermodynamic phase structure of Lifshitz and HSV black holes utilizing Lyapunov exponent. The article is organized as follows: In the next section (Section \ref{sec_HSV_solution_and_thermo}), we provide a quick and concise introduction to constructing a generic HSV gravity model, encompassing the Lifshitz spacetime. In the same section, we outline the critical phenomena of the spacetime in the canonical ensemble. Subsequently, in Section \ref{sec_geodesic}, we introduce the geodesic motion of massive and massless particles in connection with the Lyapunov exponent. In Section \ref{sec_lyap}, we investigate the phase structure of the black hole spacetime using the Lyapunov exponent. Finally, we conclude our findings in Section \ref{sec_discussion} with a discussion.

\section{Hyperscaling violating black hole: Solution and thermodynamics} \label{sec_HSV_solution_and_thermo}

In this section, in line with the reference \cite{Pedraza:2018eey}, we introduce the model under consideration along with its thermodynamic details. In \cite{Pedraza:2018eey}, the authors introduced an electrically charged black brane which is characterised by arbitrary Lifshitz exponent $z$ and HSV parameter $\theta$ by using a generalized EMD action. The resulting solutions includes novel features, such as spherical and hyperbolic horizon topologies, in addition to the planar horizons identified in Ref. \cite{Alishahiha:2012qu}. As particular cases, these new solutions incorporate the spherical Lifshitz black holes previously identified in \cite{Tarrio:2011de}. It is noteworthy that the admissible values for $z$ and $\theta$ are constrained by the null energy condition (NEC). The occurrence of phase transitions is exclusive to spherical black holes characterized by $1 \leq z \leq 2$, with no restrictions on $\theta$.

In this section, we present the essential equations required for the subsequent calculations. While the equation set may seem intricate, we include them for the sake of self-containment in the article. For a more detailed discussion on the model, we direct readers to Ref. \cite{Pedraza:2018eey} and the references therein.

\subsection{The black hole model and solutions}
\label{setup}
The theory describing the generic HSV black hole spacetime is a modification of the standard EMD theory \cite{Dong:2012se}. This modified theory introduces two additional vector fields, $H$ and $K$, where $H$ supports non-trivial topology, and $K$ supports states with finite charge density. The action is given by:
\begin{equation}
\label{action1}
S=-\frac1{16\pi G}\int d^{d+2}x\sqrt{-g}\left[R-\frac12(\nabla_\mu\phi)^2+V(\phi)-\frac14 X(\phi)F^2-\frac14 Y(\phi)H^2-\frac14 Z(\phi)K^2
	\right]\,,
\end{equation}
where the fields are associated with the corresponding gauge potentials as $F=dA$, $H=dB,$ and $K=dC$. The potential and dilaton couplings are chosen as:
\begin{equation}
V=V_0 e^{\lambda_0 \phi},\;\;X=X_0 e^{\lambda_1\phi},\;\;Y=Y_0 e^{\lambda_2\phi},\;\;Z=Z_0 e^{\lambda_3\phi}\,,
\end{equation}
with arbitrary constants $V_0$, $X_0$, $Y_0$, $Z_0$, and $\lambda_i$. The positive constants $X_{0}$, $Y_{0}$, and $Z_{0}$ are the magnitude of the coupling between the gauge fields and gravity. In natural units, the solution to the field equations is given by:
\begin{equation} \label{ansatz}
\begin{aligned}
  ds^2&=\left(\frac{r}{r_F}\right)^{  -2 \theta /d   }\left(
-\left(\frac r\ell\right)^{2z}f(r) dt^2+\frac{\ell^2}{f(r)r^2}dr^2+r^2 d\Omega ^2_{k,d}
\right) \,,     \\
A&=a(r) dt\,,\qquad B=b(r) dt\,,\qquad C=c(r) dt\,,\qquad \phi=\phi(r)\,.
\end{aligned}
\end{equation}
The parameter $k$ takes on values of $-1, 0, 1$, corresponding to the hyperboloid, planar, or spherical topology for the black hole horizon, where
\begin{equation}
\begin{split}
	d\Omega^2_{k=1,d}
	=
	d\chi_{0}^{2}
	+\sin(\chi_{0})^{2}d\chi_{1}^{2}
	+
	\dots
	+
	\sin(\chi_{0})^{2}\cdots \sin(\chi_{d-2})^{2}d\chi_{d-1}^{2}
	\,,\\
	d\Omega^2_{k=0,d}
	=
	\frac{d\vec{x}_{d}^{2}}{\ell^{2}}	
	\,, \qquad
	d\Omega^2_{k=-1,d}
	=
	d\chi_{0}^{2}
	+
	\sinh(\chi_{0})^{2}d
	\Omega^{2}_{k=1,d-1}
	\,,\qquad
\end{split}
\end{equation}
the angles $\chi_{i}$ represent angular coordinates. The constant $\ell$ is the generalization of the AdS radius. The parameter $z$ is the Lifshitz dynamical exponent and $\theta$ is the HSV exponent, both corresponding to the symmetries of the underlying theory. At $r\to \infty$, we anticipate $f(r)$ to approach unity. In this limit, the equation (\ref{ansatz}) stands as a comprehensive metric which maintains covariance under the scale transformations outlined in Eq. \ref{HSV_trans}. Within the framework of this model (Eq. 2.1), the introduction of $F$ supports the Lifshitz asymptotics of the geometry, $H$ contributes to the topology of internal space, and $K$ corresponds to solutions involving electric charge. The scalar potential $V(\phi)$ plays a pivotal role in facilitating the HSV factor of the solution.

The solution to the field equations are obtained in terms of the constants $z$, $\theta$, and $k$, which reads,
 \begin{equation}  \label{blackeningspherical}
 \begin{aligned}
\phi=&\phi_0+\gamma\log r,  \\
F=&-\rho _1 e^{-\lambda _1 \phi(r) }r^{-\frac{2 \theta }{d}-d+\theta +z-1}dtdr,\\
H=&-\rho_2e^{-\lambda _2 \phi(r) }r^{-\frac{2 \theta }{d}-d+\theta +z-1}dtdr,\\
K=&-\rho_3e^{-\lambda _3 \phi(r) }r^{-\frac{2 \theta }{d}-d+\theta +z-1}dtdr,\\
f =&1     +k\frac{(d-1)^2  }{  (d-\theta +z-2)^2} \frac{\ell^2}{r^2} -  \frac{m}{r^{ d-\theta +z} } +  \frac{ q^2}{ r^{2 (d- \theta + z-1)}}  
\end{aligned}
\end{equation}
Here, $\gamma\equiv\sqrt{2\left(d-\theta\right)\left(z-1-\theta/d\right)}$, and the remaining quantities are defined as:
\begin{equation}
\begin{aligned}
\lambda_0 =& \frac{2 \theta  }{\gamma d } , \;\;
\lambda _1 = -\frac{2 \left(d- \theta +\theta /d \right)}{ \gamma  },\;\;  \lambda _2= -\frac{2 (d-1) (d-\theta )}{ \gamma  d}, \;\;    \lambda _3= \frac{\gamma }{d-\theta },  \\
V_0=& (d-\theta +z-1) (d-\theta +z) \ell^{-2} r_F^{- 2 \theta  /d} e^{-\lambda_0 \phi_0 },\\
\rho _1^2=&  2       (z-1) (d-\theta +z)  X_0^{-1}   \ell ^{-2 z} r_F^{ 2 \theta /d}   e^{  \lambda_1  \phi_0 }    ,\\
\rho _2^2=&2  k \frac{ (d-1)   (d (z-1)-\theta )    }{d-\theta +z-2} Y_0^{-1}  \ell ^{2(1-z)} r_F^{ 2 \theta /d} e^{\lambda_2 \phi_0 }   , \label{rho2}\\
\rho_3^2=&  2  q^2 (d-\theta) (d-\theta + z-2) Z_0^{-1}  \ell^{-2z}r_F^{2 \theta/d}e^{ \lambda_3 \phi_0 } .
\end{aligned}
\end{equation}
It is important to emphasize that the parameters $m$ and $q$, related to mass and charge respectively, are embedded in  $f(r)$, and they can assume arbitrary values as long as the black hole solution exists. It is crucial to note that the validity of the solution is contingent upon the constraints $d- \theta +z -2 > 0$ and $\gamma \in \mathbb{R}$.

Now, let us delve into the exploration of the thermodynamics and critical phenomena associated with the aforementioned black hole solutions. Notably, the family of black hole solutions characterized by spherical topology (the case where $k=1$) exhibits a non-trivial phase structure. Importantly, the qualitative features of the thermodynamics in the HSV case $(\theta \neq 0)$ are similar to the charged Lifshitz black holes case $( \theta =0)$ \cite{Tarrio:2011de}. Therefore, our analysis in the following sections, particularly regarding the Lyapunov exponent, will include both the HSV and Lifshitz cases.

From the perspective of the dual theory, hyperscaling is the property that the free energy of the system scales with its naive dimension. At finite temperature, theories exhibiting hyperscaling feature an entropy density scaling with temperature as $S \sim T ^{d/z}$. However, when hyperscaling is violated, a modified relationship emerges, $S \sim T^{(d-\theta)/z}$,  suggesting that the system effectively resides in a dimension $d_{eff} = d-\theta$ \cite{Huijse:2011ef, Sachdev:2012dq}. Roughly speaking, in a theory with hyperscaling violation, the thermodynamic behavior mimics a theory with a dynamical exponent \(z\) but inhabits \(d-\theta\) dimensions. Dimensional analysis is restored in these theories because they typically involve a dimensionful scale that does not decouple in the infrared, giving rise to such behavior. One can then employ appropriate powers of this scale, denoted as \(r_F\), to restore naive dimensional analysis \footnote{The case \(\theta =d-1\) offers a promising gravitational representation of a theory with a Fermi surface in terms of its leading large \(N\) thermodynamic behavior. In this scenario, the relevant dimensionful scale is, of course, the Fermi momentum \cite{Huijse:2011ef, Ogawa:2011bz}.}.

Charged black holes exhibit a complex structure with multiple inner horizons and a distinct outer horizon (event horizon).  This event horizon, denoted as $r_h$, is determined by finding the largest positive root of the equation $f(r_h) = 0$. In addition to the horizon radius $r_h$, the black hole system is characterized by various length scales: the generalised AdS radius $\ell$, the ultra-violet scale $r_F$, the scalar amplitude $\phi _0$, and the charge parameter $q$. These parameters collectively govern the thermodynamic properties of the black hole. The mass parameter, $m$ can be written in relation to the horizon radius $r_h$ by satisfying the condition $f(r_h) = 0$, as expressed in the equation
\begin{equation}\label{massparameter}
m = r_h^{d+z-\theta} \left [ 1+ k \frac{ (d-1)^2  }{ (d-\theta +z-2)^2} \frac{\ell^{2}}{r_h^2}
+ \frac{q^2} { r_h^{2 (d-\theta  + z - 1)} } \right] \, .
\end{equation}
For the spherical black holes of interest in this article, $m$ is non-negative. The Hawking temperature $(T)$ is computed using standard Euclidean trick:
\begin{equation}\label{HawkT}
T = \frac{1}{4\pi} \left ( \frac{r_h}{\ell} \right)^{z+1} \big | f ' (r_h) \big |
\,.
\end{equation}
The absence of conformal factor, which encompasses the ultraviolet scale $r_F$, in this formula is related to the conformal invariant of Hawking temperature \cite{Jacobson:1993pf}. Substituting the expressions for $f(r)$ from Eq. \ref{blackeningspherical} for the blackening factor and (Eq. \ref{massparameter}) for the mass parameter into the temperature formula yields
\begin{equation}   \label{temperature}
T = \frac{r_h^{z}}{4 \pi \ell^{z+1}}  \left [(d-\theta +z)
 +k \frac{(d-1)^2  }{  (d-\theta +z-2)}  \frac{\ell^2}{r_h^2}
 -  \frac{ (d-\theta + z -2)  q^2 }{  r_h^{2 ( d- \theta + z-1)} }
   \right] \, .
\end{equation}

The entropy is determined by the area law,
\begin{equation}
S = \frac{\omega_{k,d}}{4 G} r_h^{d-\theta} r_F^\theta \,
\,.
\end{equation}
where $\omega_{k,d}$ defined by the unit metric $d \Omega_{k,d}^2$ represents the volume of the space. Notably, the entropy remains independent of $z$ but explicitly relies on $\theta$. Moreover, extremality is attained when the temperature tends to zero. This occurs when the charge parameter is given by
\begin{equation} \label{extremalcharge}
 q^2_{ext} =      r_{ext}^{2(d+z-\theta -1)} \left [\frac{d-\theta+z}{d-\theta + z- 2 } + k \frac{\ell^2}{r_{ext}^2} \frac{(d-1)^2}{(d-\theta + z - 2 )^2} \right] \, ,
 \end{equation}
with $r_{\text{ext}}$ representing the horizon radius of the extremal black hole which is characterised by $f(r_{\text{ext}}) = f'(r_{\text{ext}}) = 0$. The corresponding mass parameter in terms of $r_{\text{ext}}$ yields
\begin{equation} \label{extremalmass}
 m_{ext}= 2  r_{ext}^{d-\theta +z}  \left [       \frac{d-\theta +z-1}{d-\theta +z-2} + k  \frac{\ell^2}{r_{ext}^2} \frac{  (d-1)^2    }{(d-\theta +z-2)^2}   \right]
 \,.
 \end{equation}
The extremal solution signifies the ground state in the canonical ensemble, and its finite entropy indicates a high degree of degeneracy—a well-established trait shared with charged AdS black holes \cite{Chamblin:1999tk}. The ADM mass is given by
\begin{equation} \label{sphericalmass}
 \qquad M =   \frac{\omega_{k,d}}{16 \pi G} (d- \theta) m \ell^{-z-1} r_F^\theta
\,.
\end{equation}
For $\theta=0$, this expression reduces to that of pure Lifshitz case.

The overall electric charge of the black hole is determined by the conserved charge associated with the field strength $K$:
\begin{eqnarray} \label{electriccharge}
Q \equiv Q_K &=& \frac{1}{16\pi G} \int Z(\phi ) * K = \frac{\omega_{k,d}}{16\pi G} Z_0  \rho_3 \ell^{z-1} r_F^{\theta- 2 \theta/d} \,\\ 
&=& \frac{\omega_{k,d}}{16 \pi G}  \sqrt{2 Z_0 (d-\theta)(d-\theta + z-2)} \, q  \, \ell^{-1} r_F^{\theta -  \theta/d} e^{\lambda_3 \phi_0 /2} \nonumber
\end{eqnarray}

It is noteworthy that analogous expressions exist for the remaining two conserved charges $Q_F$ and $Q_H$ in relation to $\rho_1$ and $\rho_2$. Nevertheless, these two charges lack a direct thermodynamic interpretation. \footnote{The thermodynamics does not involve the gauge potentials A and B. These fields are included in the theory only to maintain the required structure and geometry of the spacetime. 
However, modifying the charges associated with these potentials, namely $Q_F$ and $Q_H$, would entail altering the symmetries and geometry of the boundary field theory, significantly impacting the holographic interpretation. We note that this could serve as motivation to initiate the exploration of the holographic thermodynamics (which has gained significant interest in the community recently \cite{Ahmed:2023snm}) of HSV spacetime.}

The verification of the first law of thermodynamics is straightforward using the defined  thermodynamic quantities. In the canonical ensemble, a comparison to the extremal case, rather than the thermal case, necessitates a modified form of the first law:
\begin{equation}
d \hat{M} = T dS + \hat{\Phi} dQ \, ,\label{first_lawB}
\end{equation}
Here, the adjusted mass and electric potential are articulated as:
\begin{eqnarray} 
&& \hat{M} = M - M_{ext} = \frac{\omega_{k,d}}{16 \pi G} (d-\theta) (m - m_{ext}) \ell^{-z-1} r_F^\theta   \, , \label{masscanonical} \\
&& \hat{\Phi} \,\, = \Phi - \Phi_{ext} = \frac{q}{c} \left (  \frac{1}{r_h^{d- \theta + z-2}} -  \frac{1}{r_{ext}^{d- \theta + z-2}}   \right) \, . \label{potential}
\end{eqnarray}
It is essential to note that each of the aforementioned thermodynamic quantities, along with the specific expressions of the first laws, can be obtained by using the Euclidean method (refer to Appendix A in \cite{Pedraza:2018eey}).


\subsection{Thermodynamics in canonical ensemble }
\label{sec:canonical}

Our objective is to explore the black hole phase structure within the canonical ensemble, where the charge $Q$ is held constant, while the potential at infinity $\Phi$ is permitted to change. The free energy can be obtained by the standard formula as,
\begin{equation}
\begin{aligned}
 F & \quad = \quad      \hat M - T S   \\
 & \quad= \quad  \frac{\omega_{k,d}}{16 \pi G} \ell^{-z-1} r_F^\theta \Big [    - m_{ext} (d-\theta) - z r_h^{d-\theta +z}  + k  \frac{(d-1)^2   (2-z)}{  (d-\theta +z-2)^2} \ell^2 r_h^{d-\theta+z-2}   \\
 &\qquad \quad + (2 d -2\theta+z -2 ) q^2    r_h^{ -(   d -\theta + z-2) } \Big] \, ,
 \end{aligned}
 \end{equation}
where $m_{ext}$ is given by (\ref{extremalmass}). Although our primary interest is in the spherical black hole case $(k=1)$, this result remains applicable for each $k$ as the extremal black hole consistently represents the reference background in the canonical ensemble. Within the canonical ensemble, both solutions possess an identical charge parameter $q=q_{ext}$, defined by (Eq. \ref{extremalcharge}). The horizon radius $r_h$ implicitly depends on $T$ and $q$ through (Eq. \ref{temperature}), allowing $F$ to be regarded as a function of $T$ and $q$.

In reference \cite{Pedraza:2018eey}, it was discovered through numerical inspection that for planar and hyperbolic black holes, the thermodynamic potential $F<0$. This explains the absence of a phase transition for these kind of black holes for all possible values of $\theta$ and $z$.  
However, for spherical black holes, there exist a phase transition which closely resembles that of charged AdS black holes revealed in \cite{Chamblin:1999tk,Chamblin:1999hg}. The phase diagram for spherical black holes exhibits a consistent qualitative pattern for every $\theta$, yet it varies for different values of $z$. This is the reason we confine our focus to spherical HSV solutions in this article.

\begin{figure}[ptb]
  \includegraphics[width=\textwidth]{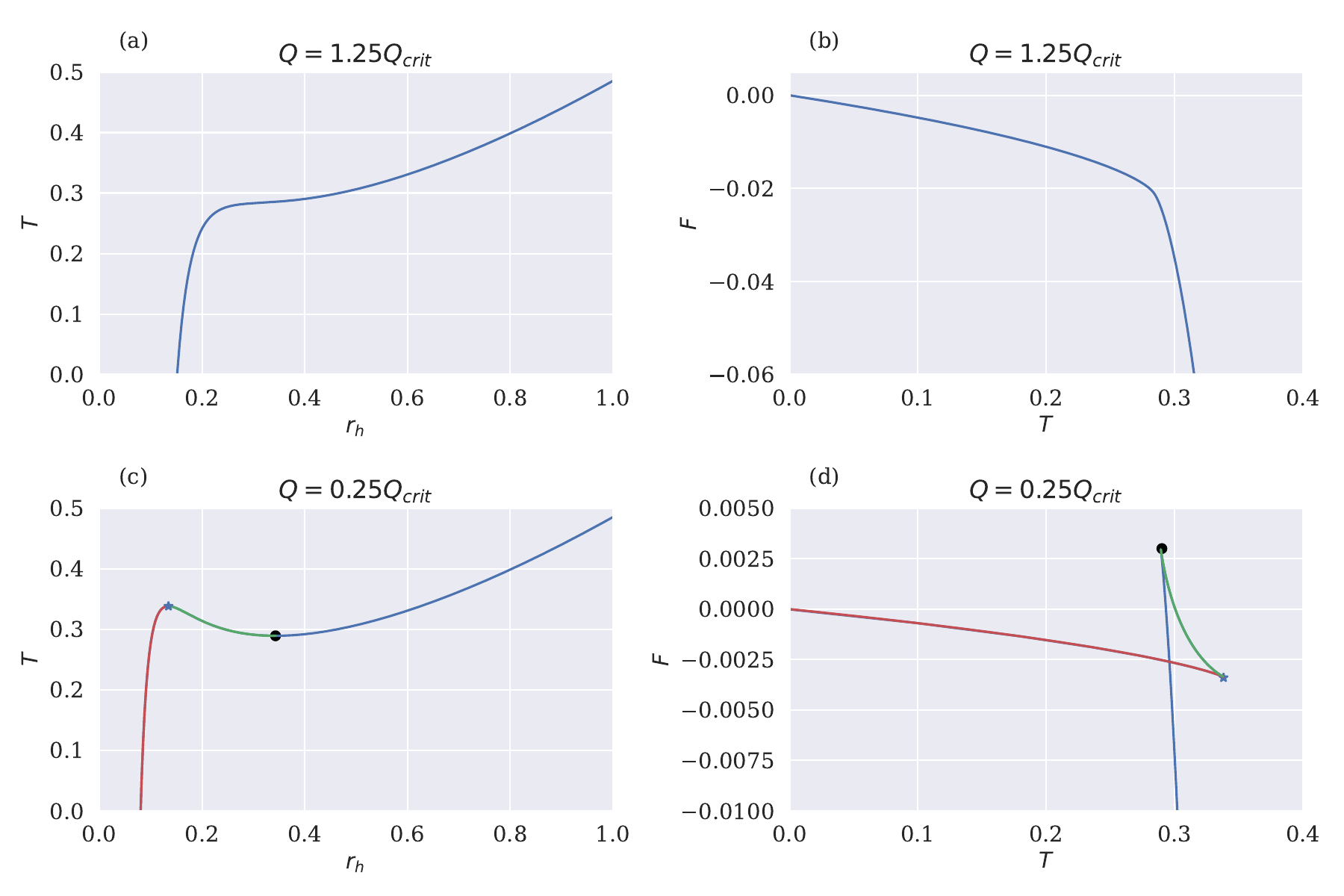}
  \caption{Hawking temperature vs. horizon radius and free energy vs. Hawking temperature depicted for two distinct charge regimes: $Q=1.25  Q_{\text{crit}}> Q_{\text{crit}}$ (top row) and $Q=0.25  Q_{\text{crit}}< Q_{\text{crit}}$ (bottom row). The figures are generated with fixed spacetime parameters, including $(\theta=0, k = r_F = Z_0 = \Phi_0 = 1, d = 3, \Omega_{kd} = 16 \pi G, z=3/2)$. The qualitative behavior of these figures remain same for all allowed values of $\theta$ and $z$. }%
  \label{tr_and_ft}%
\end{figure}


Figure \ref{tr_and_ft} showcases $T-r_h$ and $F-T$ plots for two distinct charge values, highlighting distinctive patterns for values both above and below a critical threshold, identified as $Q_{\text{crit}}$. (The critical value is computed shortly). For charges above critical value ($Q > Q_{\text{crit}}$), the temperature $T$ exhibits an injective relationship with $r_h$, and $F$ remains always negative, which underscores the dominance of charged black hole solutions in the phase portrait. Conversely, for small charges ($Q < Q_{\text{crit}}$), a temperature range emerges where three distinct branches of black hole solutions coexist (indicated by the blue star and black circle). Here, the thermodynamic potential $F$ showcases the characteristic "swallowtail" pattern observed in charged AdS black holes \cite{Chamblin:1999tk,Chamblin:1999hg}.

In the $F-T$ plot for $Q < Q_{\text{crit}}$, three distinct branches are observed. The first branch, referred to as the small black hole (SBH) branch, originates from the origin and terminates at the blue star. The second branch, known as the intermediate black hole (IBH) branch, extends between the blue star and the black circle. The third branch, representing the large black hole (LBH) branch, exhibits a cusp with the second branch at the black circle and continues downward as temperatures increase. The occurrence of a first-order phase transition between a SBH nad LBH phase is marked precisely when the SBH and LBH branches intersect in the $(F, T)$ diagram. The IBH branch, characterized by a negative heat capacity at constant charge $C_Q$, is thermodynamically unstable and does not influence the phase diagram. As $Q$ approaches 
$Q_{\text{crit}}$, the IBH branch gradually diminishes, and the SBH and LBH branches eventually merge. At $Q = Q_{\text{crit}}$, a second-order phase transition between SBH and LBH phases persists, designating this point in the phase portrait as a genuine critical point. Beyond $Q_{\text{crit}}$, no further phase transitions occur.

Analyzing the temperature in relation to the horizon radius allows for the determination of the critical charge $Q_{\text{crit}}$. Notably, for $Q < Q_{\text{crit}}$, the temperature displays two turning points, while for $Q > Q_{\text{crit}}$, no turning points are observed. At the critical charge $Q = Q_{\text{crit}}$, the temperature exhibits an inflection point with the following conditions:
\begin{equation}
 \frac{\partial T}{\partial r_h} = 0 \quad \text{and} \quad \frac{\partial^2 T}{\partial r_h^2}=0   \qquad \text{at} \qquad r_h = r_{crit}\,,\quad q= q_{crit} \, .
 \end{equation}
Solving these equations yields the critical values:
 \begin{equation}  \label{critical}
 r_{crit}^2 = k \frac{   (d-1)^2  (2-z)\ell^2 }{z (d-\theta +z-1) (d-\theta
   +z)} \quad \text{and} \, , \quad q_{crit}^2 = \frac{z  (d-\theta +z) r_{ crit}^{2(d-\theta +z-1)}  }{(d-\theta +z-2)^2 (2 d -2\theta+z -2)}\,.
 \end{equation}
For $(\theta = 0, z = 1)$, these results align with findings in \cite{Chamblin:1999tk}, and for arbitrary $z$ values, they agree with results in \cite{Tarrio:2011de}. The value of critical temperature is given by:
\begin{eqnarray} \label{criticaltemp}
 T_{crit} =   \frac{ (d-\theta +z-1) (d-\theta +z)}{\pi  (2-z) (2 d-2\theta+z -2)} \frac{r_{crit}^z}{\ell^{z+1}}
 \,.
 \end{eqnarray}

Analyzing these critical quantities elucidates a crucial distinction: for $z>2$ or $k\neq1$, the absence of a critical temperature is evident. This agrees with the absence of phase transitions in hyperbolic and planar HSV black holes within the canonical ensemble. Additionally, the critical point persists for all physically plausible values of $\theta$ that adhere to the NEC. The positive nature of $T_{crit}$ stems directly from the constraints imposed by the NEC. Therefore, while the critical point value may vary depending on $\theta$, the phase structure remains qualitatively the same for all $\theta$. Furthermore, the occurrence of phase transitions depends on $z$, with the critical point existing only for $1\leq z<2$. In the case of $z=2$, the Hawking-Page phase transition remains at $Q=0$, while for $z>2$, the black hole solution dominates the whole phase portrait excluding the origin. These findings are same as obtained for pure Lifshitz case in \cite{Tarrio:2011de}.

In summary, HSV black holes in the canonical ensemble exhibit a phase structure similar to charged AdS black holes, drawing parallels with the vdW liquid-gas system. Both systems feature a line of first-order phase transitions between distinct phases, culminating in a critical point. This analogy was explored further in Ref. \cite{Pedraza:2018eey}, revealing that the critical exponents characterizing HSV black holes align perfectly with those of the vdW fluid. Intriguingly, these critical exponents remain invariant, irrespective of the values of $z$ or $\theta$. This robust universality, reminiscent of mean-field theory expectations, underscores the insensitivity of the system to microscopic details. Subsequent sections will delve into the behaviour of Lyapunov exponent in HSV black holes, reinforcing its agreement with these universal characteristics.

In the rest of the paper we work with the rescaled thermodynamic quantities for convinience. By dimensional analysis, we find that the physical quantities scale as powers of $l$, 
\begin{equation}
    \tilde Q = Q/l^{(d- \theta + z-1)} \qquad \tilde r_h= r_h/l \qquad  \tilde T= T l \qquad \tilde M =M/l^{(d-1)} \qquad \tilde r=r/l
\end{equation}
where the tildes denote dimensionless quantities \footnote{However, we use tilded quantities and non tilded quantities alternatively in the discusssion.}.

\section{Geodesic motion and Lyapunov Exponents} \label{sec_geodesic}

In this section, we concisely outline the calculation establishing the relationship between the principal Lyapunov exponent ($\lambda$) for unstable orbits and the effective potential in the radial motion of both massless and massive particles. This derivation takes a more generalized form, specifically applicable to HSV spacetimes where $g_{tt} \neq 1/g_{rr}$. Our approach follows the narrative presented in Ref. \cite{Cornish:2003ig, Cardoso:2008bp} \footnote{Ref. \cite{Cornish:2003ig} provides the initial steps of the derivation of Lyapunov exponent, emphasizing the selection of a well-defined time coordinate. The subsequent steps are detailed in Ref. \cite{Cardoso:2008bp}, introducing a concise formula expressing the principal Lyapunov exponent $\lambda$ in terms of the second derivative of the effective potential for radial motion. It is crucial to note the metric signature difference between Ref. \cite{Cardoso:2008bp} and our approach. Our metric signature is $(-1, +1,+1,+1)$, while in their paper, it is $(+1, -1,-1,-1)$. This leads to an overall negative sign in front of $V_r^{\prime \prime}$ in the expression for the Lyapunov exponent.}. We consider particle motion in the background of a static, stationary, and spherically symmetric spacetime, characterized by the metric (for our application, the HSV metric is given by Eq. \ref{ansatz}),
\begin{equation}
g_{\mu \nu} dx^{\mu} dx^{\nu}=g_{tt} dt^2+g_{rr} dr^2+g_{\theta \theta} d\theta^2+g_{\varphi \varphi} d\varphi ^2
\end{equation}

We narrow our focus to unstable circular geodesics lying on the equatorial hyperplane with $\theta = \pi/2$, reducing the problem to a system with a two-dimensional phase space. The Lagrangian governing this scenario is given by
\begin{gather}
\begin{split}
  2\mathcal{L}&=g_{\mu \nu}  \dot x^\mu \dot x^\nu \\
  &=g_{tt}  \dot{t}^2+g_{rr} \dot
  {r}^{2}+g_{\varphi \varphi} \dot{\varphi}^{2}.
\end{split}
\end{gather}
Here, dots and primes represent derivatives with respect to the proper time $\tau$ and the radial coordinate $r$, respectively. The generalized momenta $p_\mu =\partial \mathcal{L}/\partial x^\mu$ are expressed as
\begin{gather}
\begin{split}
p_t&=g_{tt}\dot t=- E_n =const,\\
p_\varphi&=g_{\varphi \varphi}\dot\varphi=L =const,\\
p_r&=g_{rr}\dot r ,
\end{split}
\end{gather}
The inversion of these momenta for $\dot \varphi$ and $\dot t$ yields
\begin{equation} \label{tdot_and_phidot}
    \dot \varphi = \frac{L}{g_{\varphi \varphi}} \hspace{1cm} \dot t = -\frac{E_n}{g_{tt}}
\end{equation}
The Hamiltonian is expressed as
\begin{gather}
\begin{split}
 2\mathcal{H}&=2(p_t \dot t +p_\varphi \dot \varphi +p_r \dot r -\mathcal{L})\\
 &=g_{tt}  \dot{t}^2+g_{rr} \dot
  {r}^{2}+g_{\varphi \varphi} \dot{\varphi}^{2}\\
  &=\frac{L^2}{g_{\varphi \varphi}}+\frac{E_n^2}{g_{tt}}+g_{rr} \dot r^2=\delta _1= const .
\end{split}
\label{hamiltonian}
\end{gather}
Here, $\delta_1 = -1,\, 0$ for time-like and null geodesics, respectively. The radial motion is described by
\begin{equation}
  \dot{r}^{2}+V_{\text{eff}}\left(  r\right)  =0,
\end{equation}
where the constant $E_n$ is interpreted as the energy and the energy per unit mass for massless and massive particles, respectively. Here, we introduce the
effective potential,
\begin{equation} 
  V_{\text{eff}}\left(  r\right)  =\frac{1}{g_{rr}}\left[  \frac{L^2}{g_{\varphi \varphi}}+\frac{E_n^2}{g_{tt}}-\delta
    _{1}\right]  .
\label{veff}    
\end{equation}
In the absence of anisotropic time scaling, this expression reduces to the effective potential introduced in \cite{Cardoso:2008bp}, employed in subsequent studies like \cite{Guo:2022kio, Yang:2023hci, Lyu:2023sih} for exploring the phase structure of black holes in Anti-de Sitter (AdS) spacetimes.

Circular orbits, where $r$ is constant, are characterized by $V_{\text{eff}}^{\prime}(r) = 0$. Solving this equation provides the radius of the orbit. Circular orbits for which $V_{\text{eff}}^{\prime\prime}(r) < 0$ are deemed \emph{unstable}. Perturbing such an orbit with a slight increase in energy will lead it to either plunge into a black hole or diverge towards infinity. Thus, the radius of an \emph{unstable} circular geodesic is determined by
\begin{equation}
V_{\text{eff}}^{\prime}(r) = 0,\quad V_{\text{eff}}^{\prime\prime}(r) < 0.
\end{equation}
Utilizing Eqs. \ref{hamiltonian} and \ref{veff}, the Hamiltonian can be expressed in terms of the effective potential as
\begin{equation}
    2\mathcal{H} = g_{rr}  V_{\text{eff}} +\delta _1 +\frac{p_r^2}{g_{rr}}
\end{equation}
The equations of motion follow as
\begin{gather}
\begin{split}
\dot r &=\frac{\partial \mathcal{H}}{\partial p_r}=\frac{p_r}{g_{rr}}\\
\dot p_r &=-\frac{\partial \mathcal{H}}{\partial r}=-\frac{1}{2} g_{rr}^{\prime}  V_{\text{eff}}-\frac{1}{2} g_{rr}  V_{\text{eff}}^{\prime}  +\frac{1}{2} \frac{g_{rr}^{\prime}}{g_{rr}^2}p_r^2
\end{split}
\end{gather}
Linearizing the equations of motion around a circular orbit with a constant radius $r=r_c$ and using the requirements $V_{\text{eff}}=V_{\text{eff}}^\prime=0$, we obtain
\begin{gather}
\begin{split}
\delta \dot r &=\frac{\delta p_r}{g_{rr}(r_c)}\\
\delta \dot p_r &=-\frac{1}{2} g_{rr}(r_c)  V_{\text{eff}}^{\prime \prime}(r_c) \delta r
\end{split}
\end{gather}

We express the Lyapunov exponent in coordinate time, which corresponds to the time measured by an observer located far from the black hole \footnote{The utility of Lyapunov exponents, while apparent, presents discomforting limitations within the framework of general relativity. Primarily, as the Lyapunov exponents vary from orbit to orbit, they lack the broad surveying capability to capture the collective behavior of all orbits, as achieved by fractal methods. Additionally, the Lyapunov exponents measure the deviation of two neighboring orbits in time, making them heavily dependent on the chosen time coordinate. Given the relativity of time, this dependence can lead to erroneous results, including zero Lyapunov exponents for genuinely chaotic systems \cite{barrow1981chaos, barrow1982chaotic, hobill2013deterministic, Cornish:1996yg, Cornish:1996hx, Semerak:1998ah}. Notably, topological measures of chaos, such as fractals, are coordinate-invariant and unaffected by the relativism of space and time \cite{Dettmann:1994dj, Cornish:1996yg, Cornish:1996hx}. In instances where a preferred time direction exists, as in the case of a Schwarzschild black hole with a timelike Killing vector, the ambiguity of time can be mitigated. As argued in Ref. \cite{Cornish:2003ig}, from an observation position asymptotically far from the black hole, a well-defined time coordinate can be employed. As long as all timescales are conscientiously compared within the same coordinate system, meaningful comparisons can be derived.}. The Jacobian for the transformation between proper time and coordinate time is given by $dt/d\tau=\dot t$. In terms of the coordinate time $t$, the equations take the form
\begin{gather}
\frac{d  }{dt} \begin{pmatrix}  \delta r \\ \delta p_r \end{pmatrix}
 =
K \begin{pmatrix}  \delta r \\ \delta p_r \end{pmatrix}
\end{gather}
where $K$ is the linear stability matrix with components
\begin{gather}
K
 =
 \begin{pmatrix} 0 &{\displaystyle \frac{1}{\dot t\, \, g_{rr}(r_c)}} \\ \displaystyle{ -\frac{1}{2 \, \, \dot t}} \, \, g_{rr}(r_c)  V_{\text{eff}}^{\prime \prime}(r_c) & 0\end{pmatrix}
\end{gather}
The principal Lyapunov exponent $\lambda$ corresponds to the eigenvalue of the matrix $K$ \cite{Cardoso:2008bp},
\begin{equation} \label{lambda_main}
    \lambda =\sqrt{-\frac{V_{\text{eff}}^{\prime \prime}(r_c)}{2\, \dot t^2}}
\end{equation}

\subsection{Timelike geodesics (massive particles)}

\begin{figure} 
    \centering
    \includegraphics[width=\textwidth]{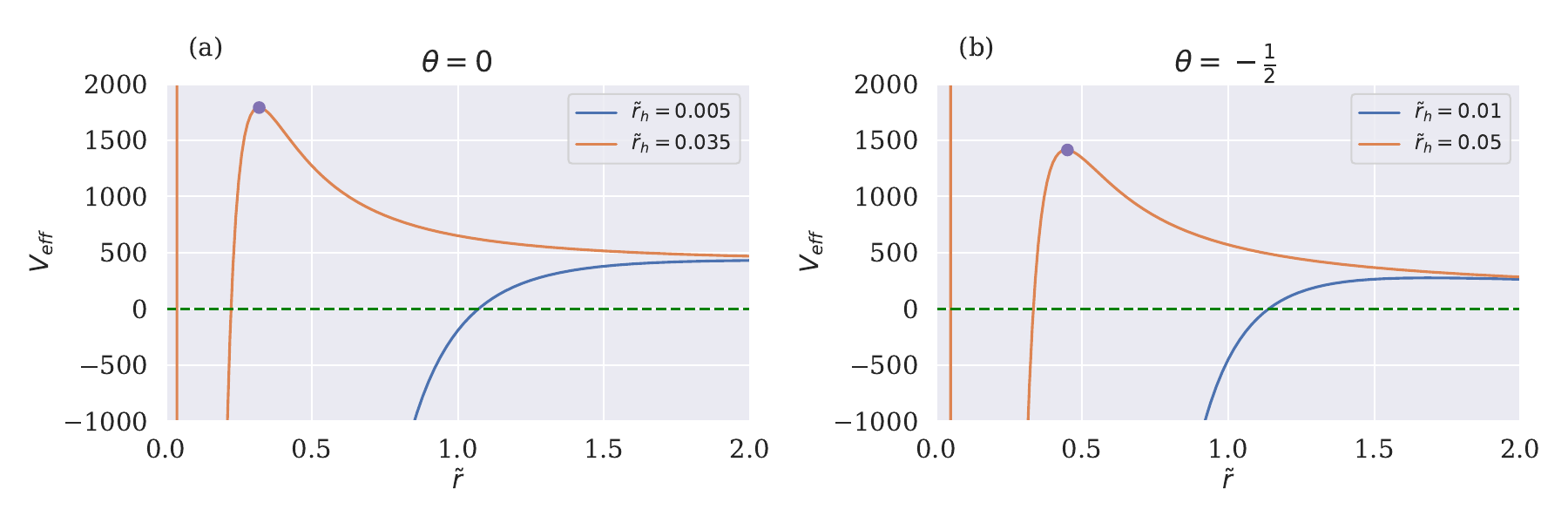}
    \caption{The effective potential $V_{\text{eff}}$ governing the motion of massive particles in HSV spacetime is depicted. Left : Lifshitz case $\theta =0 $. Right: HSV case $\theta =-1/2$. In both instances, the spacetime parameters are set to $k = r_F = Z_0 = \Phi_0 = 1$, with $d = 3$, $\Omega_{kd} = 16 \pi G$, $z = 3/2$ and $\tilde Q=0.01$. Without loss of generality, particle properties are chosen as $E_n = 1$ and $L = 20 l$. The blue dot on the plot corresponds to an unstable circular orbit. Notably, the presence of an unstable circular orbit is contingent on the size of the black hole. For black holes with substantial sizes, there is no maximum in $V_{\text{eff}}$, indicating the absence of an unstable circular orbit. }
    \label{fig_veff}
\end{figure}

In the context of HSV black hole spacetime, both stable and unstable circular geodesics are possible for massive particles. Our focus here centers on the investigation of unstable time-like circular geodesics, given their connection to the conjectured universal upper bound on Lyapunov exponents \cite{Hashimoto:2016dfz, Zhao:2018wkl}. Specifically, we delve into the Lyapunov exponent of unstable circular geodesics for massive particles possessing a defined angular momentum and energy. The effective potential governing the motion of these massive particles is expressed as follows (follows from Eq. \ref{veff}):
\begin{equation} \label{veff_massive}
  V_{\text{eff}}\left(  r\right)  =\frac{1}{g_{rr}}\left[  \frac{L^2}{g_{\varphi \varphi}}+\frac{E_n^2}{g_{tt}}+1\right]  ,
\end{equation}
Upon substituting the metric components $g_{rr}$, $g_{\varphi \varphi}$, and $g_{tt}$ from Eq. \ref{ansatz} and incorporating Eq. \ref{masscanonical} into the blackening factor $f(r)$, we find that $V_r$ is closely tied to the HSV parameter $\theta$, the Lifshitz exponent $z$, and the black hole radius $r_h$, among other spacetime parameters that are not relevant to our current analysis. The effective potential exhibits key features of timelike geodesics, revealing turning points and the locations of stable or unstable equilibria. In Figure \ref{fig_veff}, we show the effective potential energy experienced by massive particles in HSV black holes for various values of $r_h$, while maintaining constant spacetime and particle parameters. We explore two scenarios: one with a vanishing $\theta$ value, corresponding to the Lifshitz case, and another with a non-zero $\theta$ value indicative of HSV. The maxima of the effective potential signify the presence of unstable circular orbits on the equatorial plane. Remarkably, the figure demonstrates that, regardless of the $\theta$ values, if $r_h$ exceeds a critical threshold, unstable time-like circular geodesics cease to exist, implying the disappearance of their instability ($\lambda \rightarrow 0$).

Now, we aim to rewrite the expression for the Lyapunov exponent, specifically tailored for timelike geodesics (massive particles), in terms of the effective potential. The conditions for the existence of circular orbits, $V_{\text{eff}}(r_c) = V_{\text{eff}}'(r_c) = 0$, lead to the following expression for the particle's energy and angular momentum:
\begin{gather}
\begin{split} \label{constraint_timelike}
 E_n^2=\frac{g'_{\varphi \varphi}(r_c) g_{tt}^2(r_c)}{g_{\varphi \varphi}(r_c) g_{tt}'(r_c)-g'_{\varphi \varphi}(r_c) g_{tt}(r_c)}\\
 L^2=\frac{g^2_{\varphi \varphi}(r_c) g'_{tt}(r_c)}{g'_{\varphi \varphi}(r_c) g_{tt}(r_c)-g_{\varphi \varphi}(r_c) g'_{tt}(r_c)}
\end{split}
\end{gather}
The condition $E_n^2 > 0$ imposes:
\begin{equation} \label{timelike_energy_condition}
    g_{\varphi \varphi}(r_c) g_{tt}'(r_c)-g'_{\varphi \varphi}(r_c) g_{tt}(r_c)>0
\end{equation}
By employing Eq. \ref{tdot_and_phidot} and Eq. \ref{constraint_timelike} in Eq. \ref{veff_massive} and subsequently substituting it into Eq. \ref{lambda_main}, we obtain the refined expression for the Lyapunov exponent for timelike geodesics:
\begin{equation}  \label{massive_lambda}
    \lambda =\sqrt{-\frac{V^{''}_{\text{eff}}(r_c)}{2\dot t^2}}=\frac{1}{\sqrt{2}}\sqrt{\frac{g'_{\varphi \varphi}(r_c) g_{tt}(r_c)-g_{\varphi \varphi}(r_c) g'_{tt}(r_c)}{g'_{\varphi \varphi}(r_c)}V^{''}_{\text{eff}}(r_c)}
\end{equation}
It is noteworthy that, according to Eq. \ref{timelike_energy_condition}, the conditions derived from the existence of a circular orbit, combined with the requirement for an unstable orbit, $V_{\text{eff}}^{\prime\prime}(r) < 0$, ensure that the Lyapunov exponent for timelike geodesics is always real when the orbit is \emph{unstable}.

\subsection{Null like geodesics (massless particles)}

Now, let us rewrite the expression for the Lyapunov exponent concerning nulllike geodesics. For photons propagating in the background of HSV spacetime, the effective potential, as derived from Eq. \ref{veff}, is given by:
\begin{equation} \label{veff_photon}
  V_{\text{eff}}\left(  r\right)  =\frac{1}{g_{rr}}\left[  \frac{L^2}{g_{\varphi \varphi}}+\frac{E_n^2}{g_{tt}}\right]  ,
\end{equation}
The circular orbit conditions $V_{\text{eff}}(r_c)=V_{\text{eff}}'(r_c)=0$ yield:
\begin{gather} \label{photon_E_L_ratio}
\begin{split}
 \frac{E_n^2}{L^2}=-\frac{g_{tt}(r_c)}{g_{\varphi \varphi}(r_c)}
\end{split}
\end{gather}
By employing Eq. \ref{tdot_and_phidot} and Eq. \ref{photon_E_L_ratio} in Eq. \ref{veff_photon} and subsequently substituting it into Eq. \ref{lambda_main}, we deduce that the Lyapunov exponent for nulllike geodesics is:
\begin{equation} \label{photon_lambda}
    \lambda =\sqrt{-\frac{V^{''}_{\text{eff}}(r_c)}{2\dot t^2}}=\frac{1}{\sqrt{2}}\sqrt{\frac{g_{tt}(r_c) g_{\varphi \varphi}(r_c)}{L^2}V^{''}_{\text{eff}}(r_c)}
\end{equation}
Eq. \ref{photon_E_L_ratio} implies that $g_{tt}(r_c) g_{\varphi \varphi}(r_c) >0$, which, combined with the requirement for an unstable orbit, $V_{\text{eff}}^{\prime\prime}(r) < 0$, ensures that $\lambda$ is always positive for unstable photon orbits. Note that Eq. \ref{massive_lambda} and Eq. \ref{photon_lambda} reduce to their respective forms in AdS spacetime when there is no anisotropic time scaling.

\section{Lyapunov exponent and phase structure} \label{sec_lyap}

Now that we have presented all the details of thermodynamics and geodesic motion, we are prepared to explore the phase structure of the black hole using the Lyapunov exponent. Our focus lies on black hole phase transitions in the canonical ensemble, particularly the SBH to LBH phase transition, reminiscent of the vdW fluid-gas phase transition. As previously investigated, the Lyapunov exponent can characterize the phase transitions of charged black holes in AdS spacetime \cite{Guo:2022kio}. Here, we extend this method to Lifshitz and HSV black holes, the generalization of AdS spacetime. Without loss of generality, we consider a massive particle with an angular momentum of $L = 20 l$ and energy $E_n=1$ throughout the article.

\subsection{Time like geodesics (massive particles)}

To understand the relation between the Lyapunov exponent and the HSV black hole phase transition, we initially investigate the correlation between the event horizon and the Lyapunov exponent of timelike geodesics in the background metric. In order to visualize the $\lambda - r_h$ relationship, we express $\lambda$ as follows (substituting the second relation in Eq. \ref{constraint_timelike} into Eq. \ref{massive_lambda}):
\begin{equation}  \label{massive_lambda2}
    \lambda =\frac{1}{\sqrt{2}}\sqrt{\frac{g^2_{\varphi \varphi}(r_c) g'_{tt}(r_c)}{g'_{\varphi \varphi}(r_c) L^2}V^{''}_{\text{eff}}(r_c)}
\end{equation}
We observe that the Lyapunov exponent is closely related to the radius of circular unstable geodesics $r_c$, determined by $V^{\prime}_{\text{eff}}(r_c) =0$. However, there is no analytic expression for $r_c$, necessitating numerical investigation. The radius of the unstable geodesic depends on $r_h$, $Q$, and two other relevant spacetime parameters $z$ and $\theta$ (other spacetime parameters are not relevant for our study). Therefore, the value of the Lyapunov exponent $\lambda$ depends on the event horizon radius $r_h$, black hole charge $Q$, as well as $z$ and $\theta$.

\begin{figure}[ptb] 
  \centering
  \includegraphics[width=8cm]{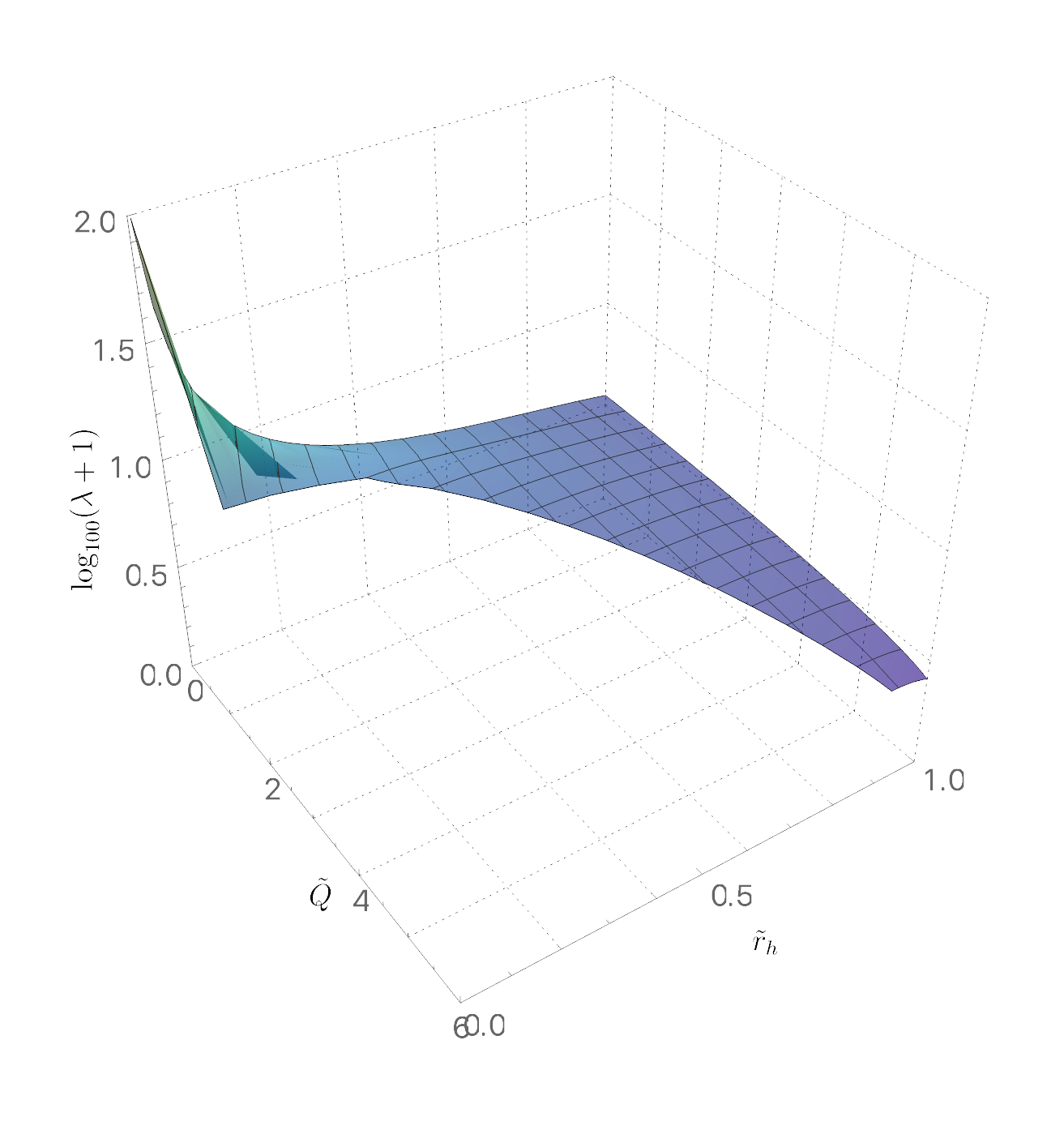} \caption{Three-dimensional plot illustrating $\log_{100} (\lambda + 1)$ as a function of $\tilde Q$ and $\tilde r_h$ for massive particles in the Lifshitz case ($\theta = 0$). The spacetime parameters are fixed with $k = r_F = Z_0 = \Phi_0 = 1$, $d = 3$, $\Omega_{kd} = 16 \pi G$, and $z = 3/2$. The exclusion of the no black hole region is enforced based on the positivity of the Hawking temperature, ensuring extremal black hole conditions.}%
  \label{density1}%
\end{figure}

\begin{figure}[t]
  \begin{center}
   \includegraphics[width=\textwidth]{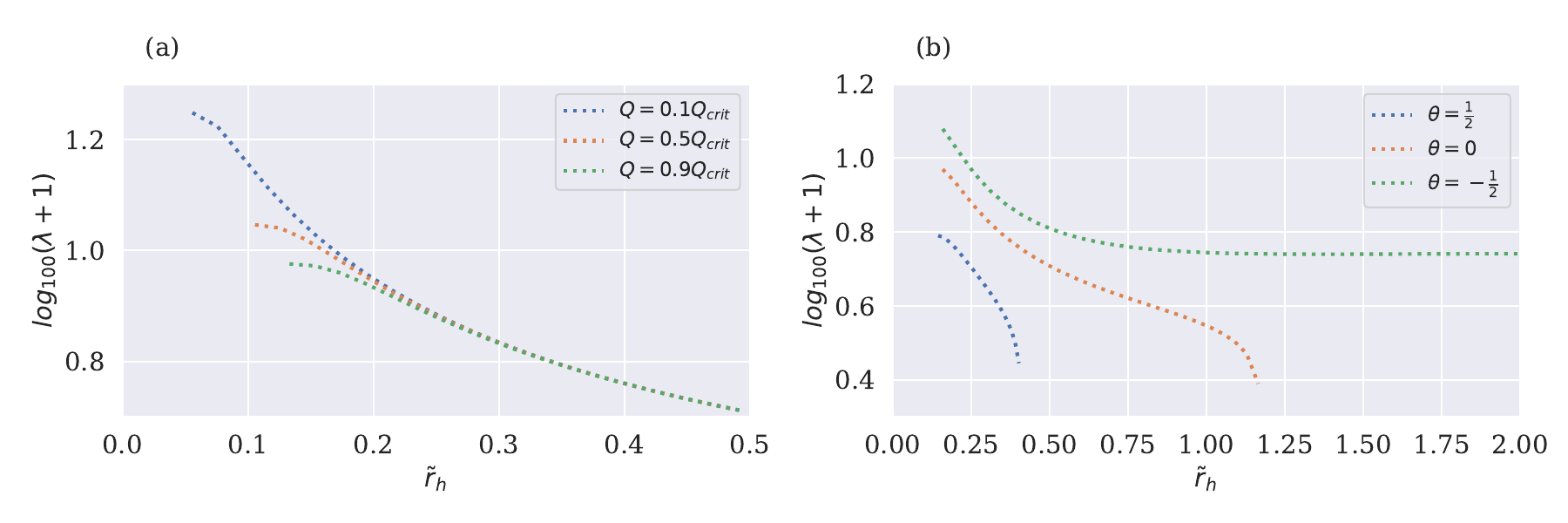}
\end{center}
\caption{The cross-section $\log_{100} (\lambda + 1) - \tilde r_h$ derived from the 3D plot with fixed $Q$ values. \emph{Left:} Illustrating the impact of charge $Q$ on $\lambda$ while maintaining other spacetime parameters constant $(k = r_F = Z_0 = \Phi_0 = 1$, $d = 3, \Omega_{kd} = 16 \pi G, z=3/2)$ in the Lifshitz case $(\theta=0)$. The selected $Q$ variation is $Q=(0.1  Q_{\text{crit}}, 0.5  Q_{\text{crit}}, 0.9  Q_{\text{crit}})$. The qualitative behavior holds for all admissible $z$ and $\theta$ values. \emph{Right:} Demonstrating the influence of HSV parameter $\theta$ on $\lambda$ while keeping other spacetime parameters fixed $(k = r_F = Z_0 = \Phi_0 = 1$, $d = 3, \Omega_{kd} = 16 \pi G, z=3/2, Q=0.9  Q_{\text{crit}})$. The chosen $\theta$ variation is $\theta =(-1/2, 0, 1/2)$. The qualitative trends remain consistent across different $\theta$ values. 
}%
  \label{2dplot1}%
\end{figure}

The 3D plot in Fig. \ref{density1} depicts $\log _{100} (\lambda +1)$ as a function of $\tilde Q$ and $\tilde r_h$, with the exclusion of regions where no black holes exist, as determined by the condition that the Hawking temperature ($T$) should satisfy $T\geq 0$. The plot illustrates that $\lambda$ diverges as $r_h$ approaches zero. Additionally, $\lambda$ tends to approach zero at specific values of $\tilde Q$ and $\tilde r_h$. This outcome aligns with the behavior observed in the RN-AdS case \cite{Guo:2022kio}. Hence, we deduce that the Lyapunov exponent associated with the geodesics of massive particles in charged black holes within HSV spacetime manifests characteristics similar to those observed in AdS spacetime. It is noteworthy that, even though we presented a 3D plot for $\theta = 0$, the observed behaviour remains consistent for non-zero $\theta$ values, as evidenced by the 2D plot (Fig. \ref{2dplot1}).

We present the cross-section of the 3D plot in Fig. \ref{2dplot1} to investigate the influence of the black hole charge ($Q$) and HSV parameter ($\theta$) on the Lyapunov exponent ($\lambda$). In the left panel, we vary $Q$ while keeping other spacetime parameters fixed (Lifshitz case, $\theta=0$; however, similar results apply to HSV scenarios). The impact of $Q$ on $\lambda$ is notably pronounced for lower values of the event horizon radius ($r_h$), and the curves converge as $r_h$ increases. The right panel of Fig. \ref{2dplot1} illustrates the effect of $\theta$ on $\lambda$ while maintaining other spacetime parameters constant. Here, the influence of $\theta$ becomes more apparent for larger values of $r_h$. In both cases, $\lambda$ tends to approach zero as $r_h$ exceeds a certain threshold, indicating the absence of unstable timelike geodesics for black holes beyond this critical $r_h$. This behavior aligns with the disappearance of the extreme point in the effective potential, as depicted in Fig. \ref{fig_veff}, where $V_{\text{eff}}^{\prime\prime} = 0 = \lambda$ at a specific $r_h$. A common feature is observed, wherein the finite domain of $r_h$ for black holes with unstable circular orbits is determined by the lower extremal black hole limit and the upper limit set by the existence of these unstable orbits. Notably, this domain exhibits significant variation with $\theta$ values, with the range expanding as $\theta$ transitions from positive to negative values, indicating an increase in unstable orbits. Throughout this study, we have chosen the convenient value $\theta = -1/2$ to exemplify hyperscaling violation.

\begin{figure}[ptb]
  \begin{center}
   \includegraphics[width=\textwidth]{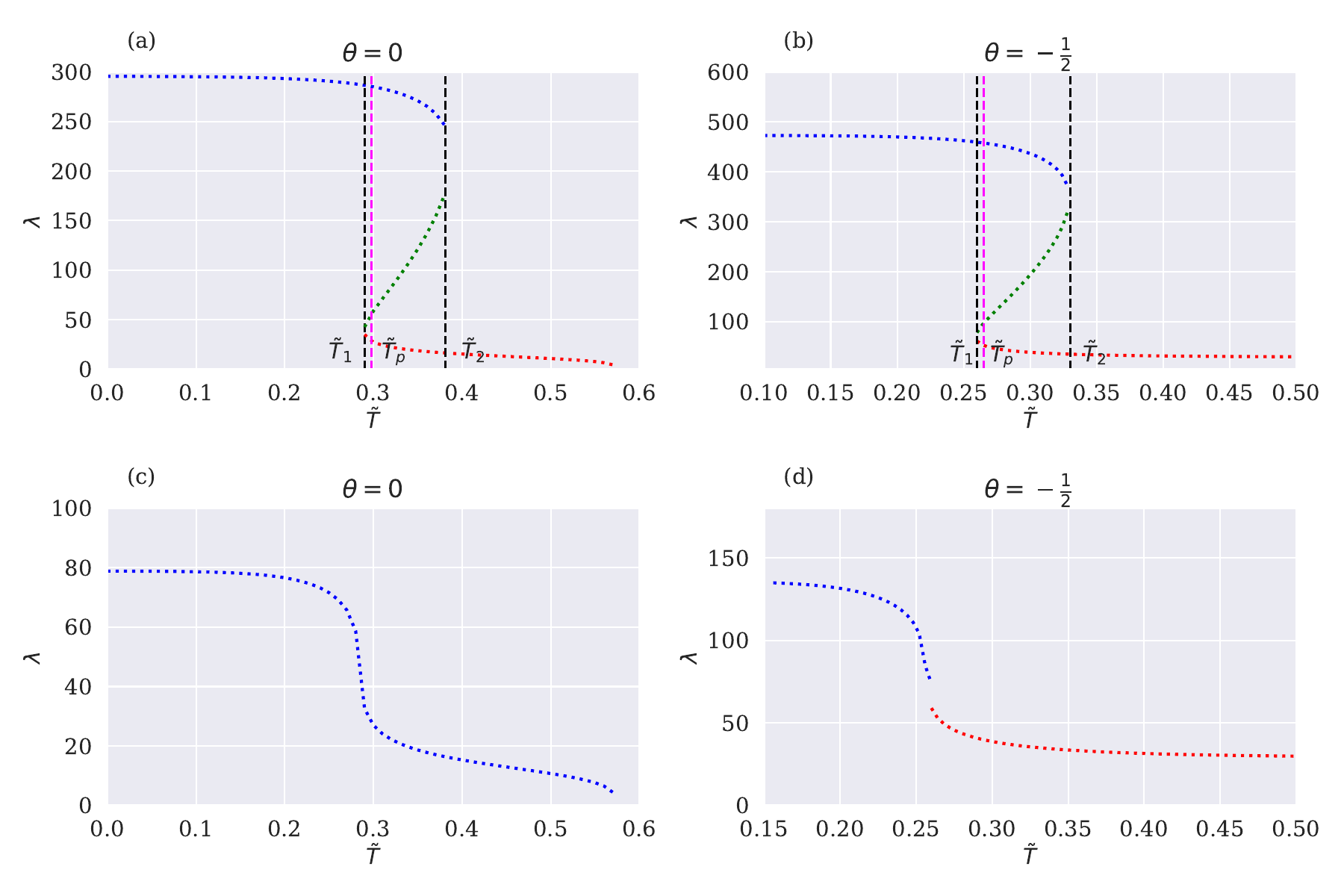}
  \end{center}
  \caption{Lyapunov exponents for massive particles plotted against the Hawking temperature for two distinct charge regimes: $Q=0.1  Q_{\text{crit}}< Q_{\text{crit}}$ (top row) and $Q=1.1  Q_{\text{crit}}> Q_{\text{crit}}$ (bottom row). The left panel corresponds to the Lifshitz case with $\theta=0$, while the right panel represents the HSV case with $\theta=-1/2$. The fixed spacetime parameters include $(k = r_F = Z_0 = \Phi_0 = 1$, $d = 3, \Omega_{kd} = 16 \pi G, z=3/2)$. The phase transition from Small Black Hole to Large Black Hole occurs at $\tilde T= \tilde T_p$.  In all cases, $\lambda$ initiates from a finite value at $\tilde T=0$ and reaches zero after a certain temperature, though this is not displayed in every plot for the sake of a clear presentation. The plots are obtained numerically, and any discontinuity near the joining points is a result of numerical instability and limitations in the number of iterations to access those points. Otherwise, the connecting plots are continuous.}%
  \label{fig:lambdaT_massive}%
\end{figure}

In our analysis, we explore the relationship between the Lyapunov exponent and the Hawking temperature. This connection is established by numerically solving the Hawking temperature Eq. \ref{temperature} to obtain $\tilde r_h(\tilde T)$ and then substituting this expression into Eqn \ref{massive_lambda} for $\lambda$. The resulting $\lambda - \tilde T$ plot is remarkably analogous to the free energy plot (Fig. \ref{tr_and_ft}). Recall that the free energy exhibits multivalued behavior below the critical point $(Q< Q_{\text{crit}})$ within a certain temperature range. This behavior signifies a first-order phase transition, where the system, initially in the SBH phase at low temperatures ($\tilde T<\tilde T_p$), transitions to the LBH phase at $ \tilde T_p$ due to a state with lower free energy. Above the critical point $Q> Q_{\text{crit}}$, the free energy is single-valued. Intriguingly, this phase transition behavior is mirrored by the Lyapunov exponent of unstable geodesics, as illustrated in Fig \ref{fig:lambdaT_massive}. We analyze two cases separately: below and above the critical point. For $Q< Q_{\text{crit}}$ (upper row in Fig \ref{fig:lambdaT_massive}), the Lyapunov exponent exhibits multivalued behavior between $\tilde T_1$ and $\tilde T_2$, corresponding to coexisting SBH, LBH, and IBH phases. In SBH, $\lambda$ initially rises to a maximum and then declines as $\tilde T$ approaches $\tilde T_2$, while for IBH and LBH, $\lambda$ increases and decreases, respectively, with rising $\tilde T$ from $\tilde T_1$. Notably, the Lyapunov exponent tends to zero as $\tilde T$ increases to a certain value. Although the qualitative behavior remains consistent, the quantitative aspects are influenced by $z$ and $\theta$. Conversely, when $Q> Q_{\text{crit}}$, $\lambda$ is single-valued, indicating a unique black hole solution. These findings suggest that $\lambda$ as a function of $\tilde T$ provides insights into the phase structure of Lifshitz and HSV black holes, with consistent behavior observed across various permissible values of $\theta$ and $z$.

Given that the $\lambda -\tilde T$ plot exhibits black hole phase transition properties similar to the behavior of free energy, it prompts the natural question of how to quantify the black hole phase transition using the Lyapunov exponent. Remarkably, the SBH-LBH phase transition can be captured by the difference in Lyapunov exponent. At the transition point $\tilde T_p$, the Lyapunov exponent takes values $\lambda _s$ for the small BH phase and $\lambda _l$ for the large BH phase. This implies a discontinuous change in the Lyapunov exponent, signifying a first-order phase transition at $\tilde T_p$. It is noteworthy that the phase transition temperature $\tilde T_p$ varies with black hole parameters, including $Q$, $z$, and $\theta$. The difference in Lyapunov exponent, $\Delta \lambda = \lambda _l - \lambda _s$, serves as an order parameter. This parameter is nonzero during the first-order phase transition and becomes zero at the critical point, highlighting its role as an order parameter. We are particularly interested in the near-critical behavior of the Lyapunov exponent, where $T_p = T_c$ and $\lambda _s = \lambda _l = \lambda _c$, marking the transition from a first-order to a second-order phase transition.

\begin{figure}[t]
  \begin{center}
   \includegraphics[width=\textwidth]{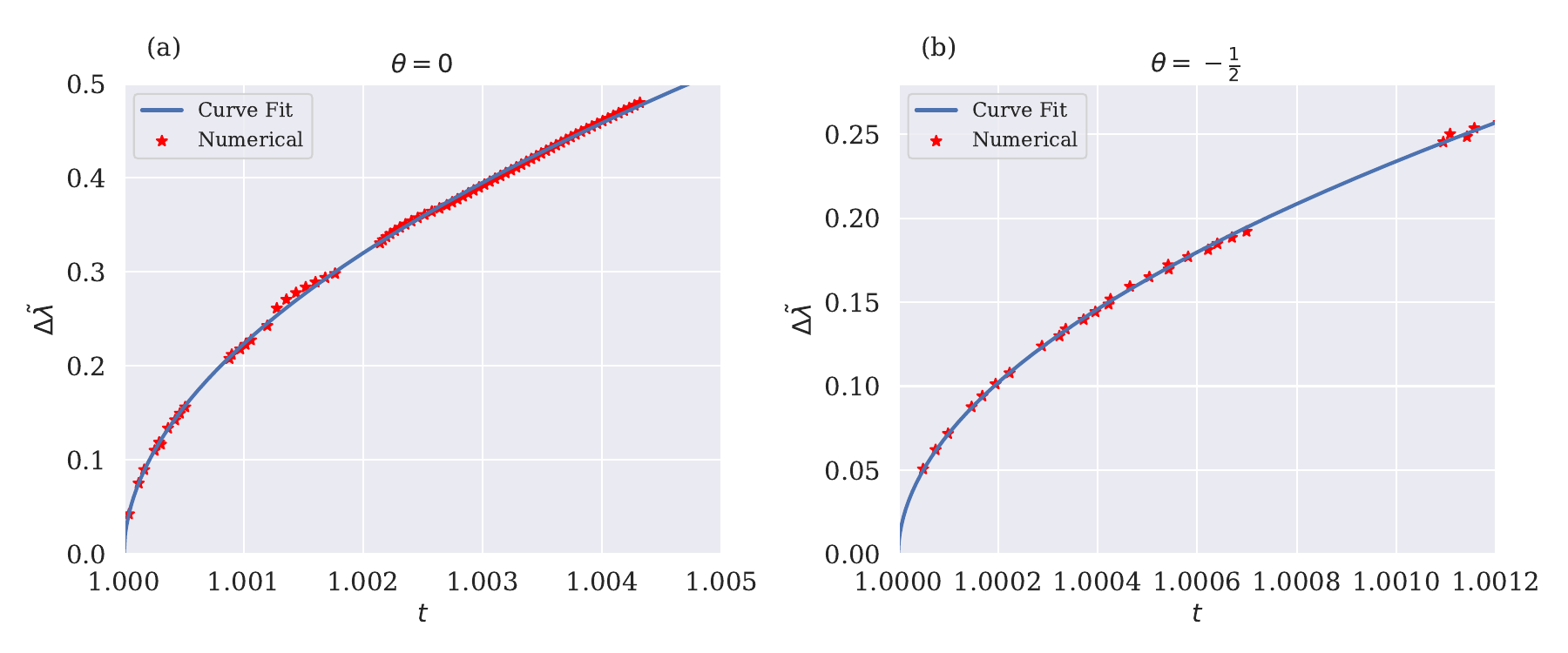}
  \end{center}
  \caption{$ \Delta \tilde \lambda$ vs $ t$ plot for massive particles. Left: Lifshitz case, $\theta = 0$. Right: HSV case, $\theta = -1/2$. The red stars represent data points obtained from numerical calculations, and the blue solid line represents the fitted curve.}%
  \label{dl_massive}%
\end{figure}

Critical exponents play a pivotal role in determining the qualitative behavior of a system in the vicinity of its critical point. To explore the critical behavior of the Lyapunov exponent for timelike geodesics around HSV black holes, we analyze the behavior of $\Delta \lambda /\lambda _c$ versus $\tilde T_p/T_c$ in Fig \ref{dl_massive} near the critical point. For simplicity, we use $\Delta \tilde \lambda$ and $ t$ to substitute $\Delta \lambda /\lambda _c$ and $\tilde T_p/T_c$ respectively. The critical exponent $\delta$ associated with the order parameter $ \Delta \tilde \lambda$ is defined as follows:
\begin{equation}
 \Delta \tilde \lambda  = \alpha |t-1|^\delta
\end{equation}
We numerically establish this relationship and find that $\delta \sim 1/2$. Specifically, for the Lifshitz case with $\theta = 0$, we obtain $\alpha = 8.04764$ and $\delta = 0.518843$, while for the HSV case with $\theta = -1/2$, we find $\alpha = 8.24289$ and $\delta = 0.515627$. Our results demonstrate that the critical exponent of $\Delta \lambda$ is identical to that of the order parameter in the vdW fluid, as predicted by mean field theory.

\subsection{Null like geodesics (massless particles)}

\begin{figure}[ptb]
  \centering
  \includegraphics[width=8cm]{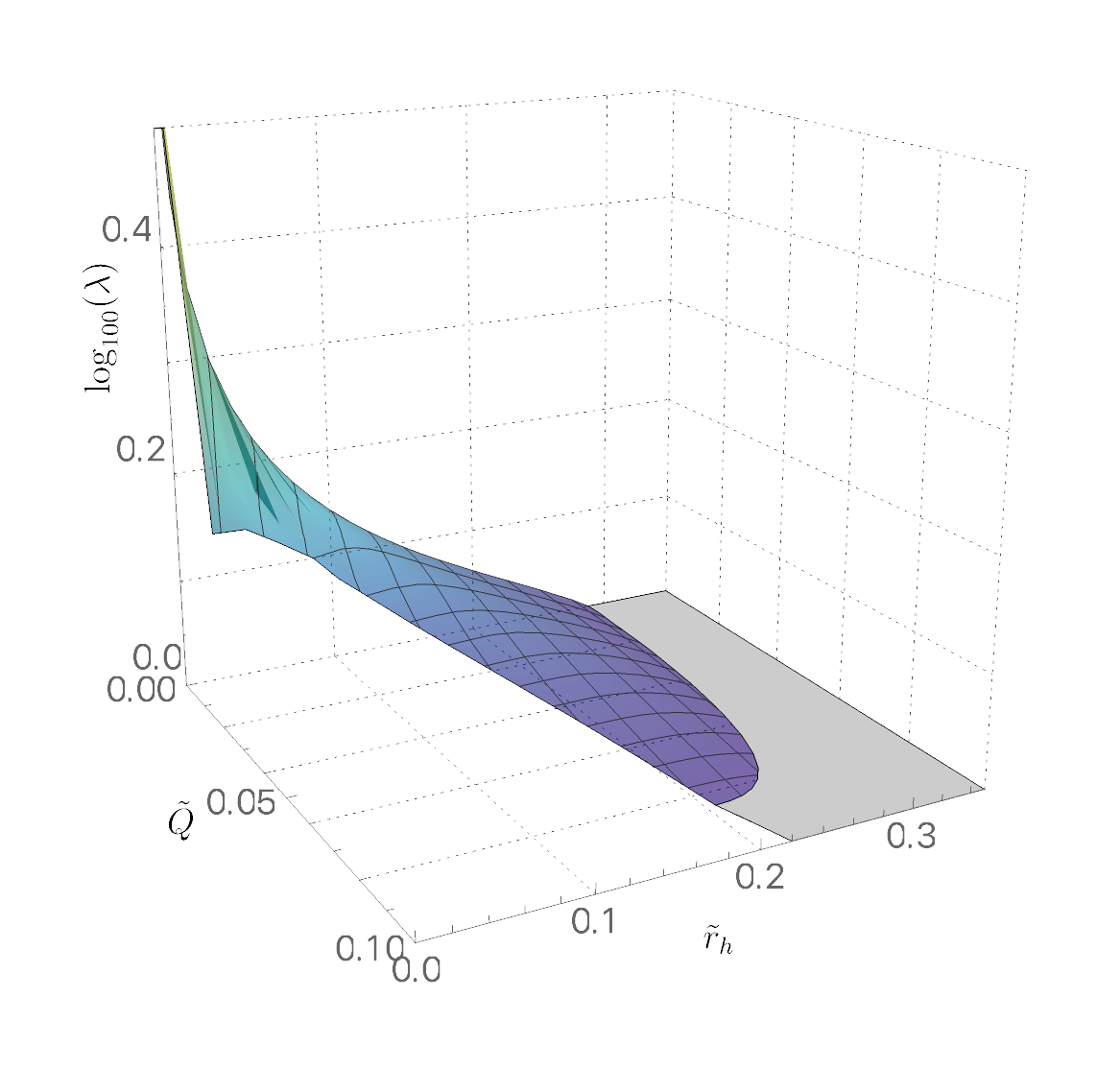} \caption{Three-dimensional plot depicting $\log_{100} (\lambda )$ as a function of $\tilde Q$ and $\tilde r_h$ for massless particles in the Lifshitz case ($\theta = 0$). The spacetime parameters are fixed with $k = r_F = Z_0 = \Phi_0 = 1$, $d = 3$, $\Omega_{kd} = 16 \pi G$, and $z = 3/2$. The exclusion of the no black hole region is enforced based on the positivity of the Hawking temperature, ensuring extremal black hole conditions. Notably, unlike AdS spacetime, the unstable null geodesics cease to exist beyond a certain $r_h$ value. The Lyapunov exponent crosses zero in the $\{Q, r_h\}$ plane in the $\lambda=0$ line and continues further, taking negative values for larger $r_h$ values. However, since we do not consider negative $\lambda$ values in our study, we truncate that portion.}%
  \label{3D_Massless}%
\end{figure}

\begin{figure}[t]
  \begin{center}
  \includegraphics[width=\textwidth]{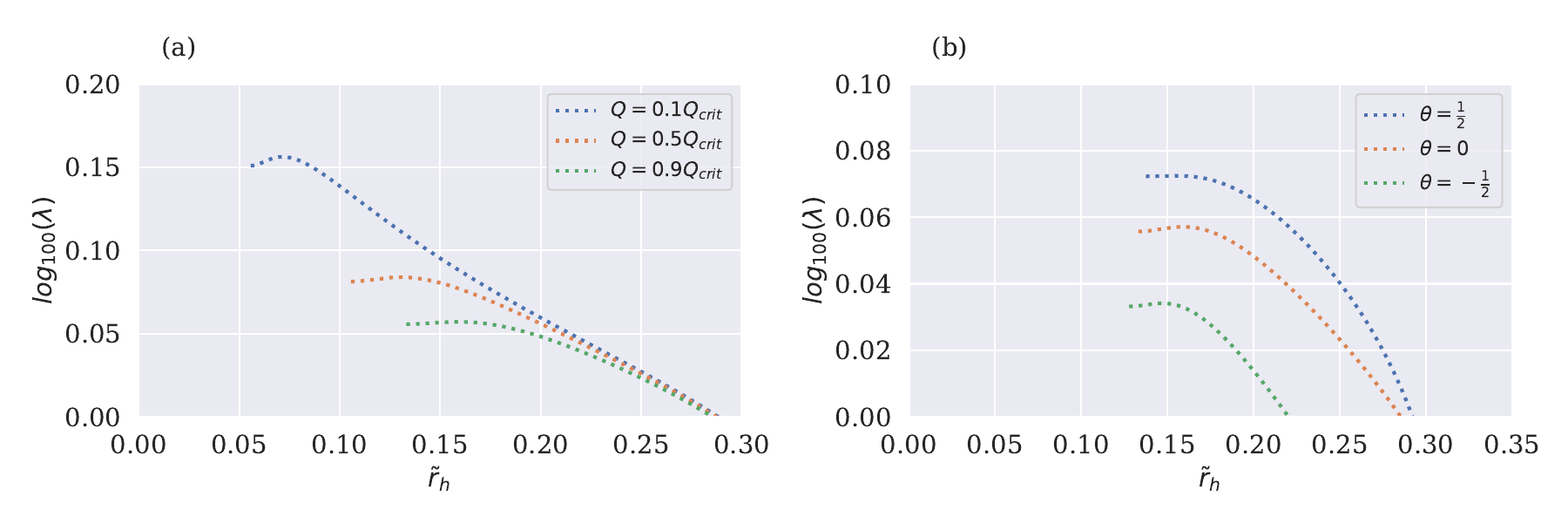}
  \end{center}
  \caption{The cross-section of $\log_{100} (\lambda ) - \tilde r_h$ derived from the 3D plot with fixed $Q$ values for massless particles. \emph{Left:} Illustrating the impact of charge $Q$ on $\lambda$ while maintaining other spacetime parameters constant ($k = r_F = Z_0 = \Phi_0 = 1$, $d = 3, \Omega_{kd} = 16 \pi G, z=3/2)$ in the Lifshitz case ($\theta=0$). The selected $Q$ variation is $Q=(0.1  Q_{\text{crit}}, 0.5  Q_{\text{crit}}, 0.9  Q_{\text{crit}})$. The qualitative behavior holds for all admissible $z$ and $\theta$ values. \emph{Right:} Demonstrating the influence of the HSV parameter $\theta$ on $\lambda$ while keeping other spacetime parameters fixed ($k = r_F = Z_0 = \Phi_0 = 1$, $d = 3, \Omega_{kd} = 16 \pi G, z=3/2, Q=0.9  Q_{\text{crit}})$. The chosen $\theta$ variation is $\theta =(-1/2, 0, 1/2)$. The qualitative trends remain consistent across different $\theta$ values. Notice that the trend is reversed in the right panel compared to the massive case (see Fig. \ref{2dplot1}). 
  }%
  \label{2d_Massless}%
\end{figure}

\begin{figure}[ptb]
  \begin{center}
  \includegraphics[width=\textwidth]{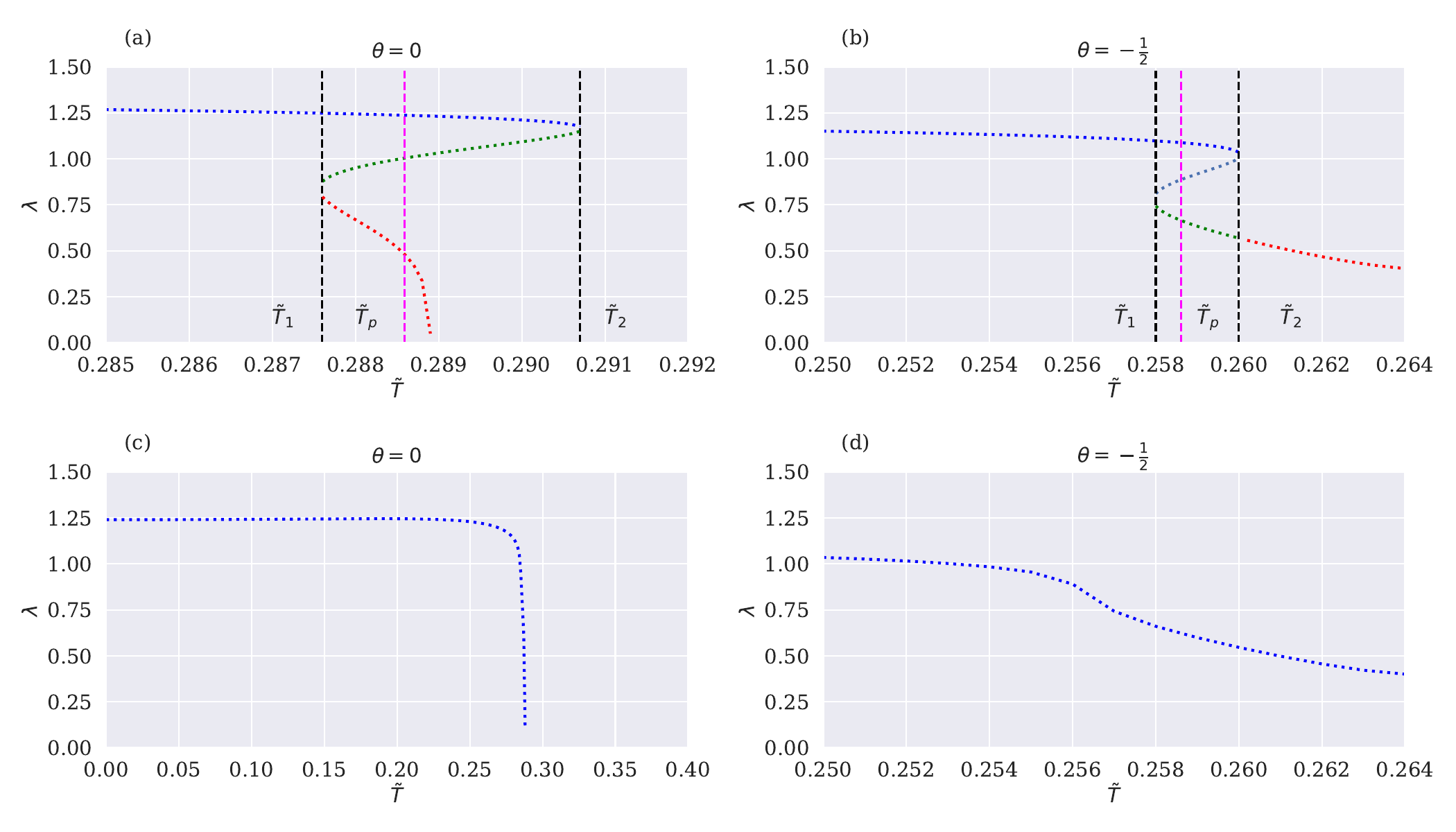}
  \end{center}
  \caption{Lyapunov exponents for massless particles plotted against the Hawking temperature for two distinct charge regimes: $Q=0.8  Q_{\text{crit}}< Q_{\text{crit}}$ (top row) and $Q=1.1  Q_{\text{crit}}> Q_{\text{crit}}$ (bottom row). The left panel corresponds to the Lifshitz case with $\theta=0$, while the right panel represents the HSV case with $\theta=-1/2$. The fixed spacetime parameters include $(k = r_F = Z_0 = \Phi_0 = 1$, $d = 3, \Omega_{kd} = 16 \pi G, z=3/2)$. The phase transition from Small Black Hole to Large Black Hole occurs at $\tilde T= \tilde T_p$. 
  }%
  \label{fig:lambdaT_massless}%
\end{figure}

\begin{figure}[t]
  \begin{center}
  \includegraphics[width=\textwidth]{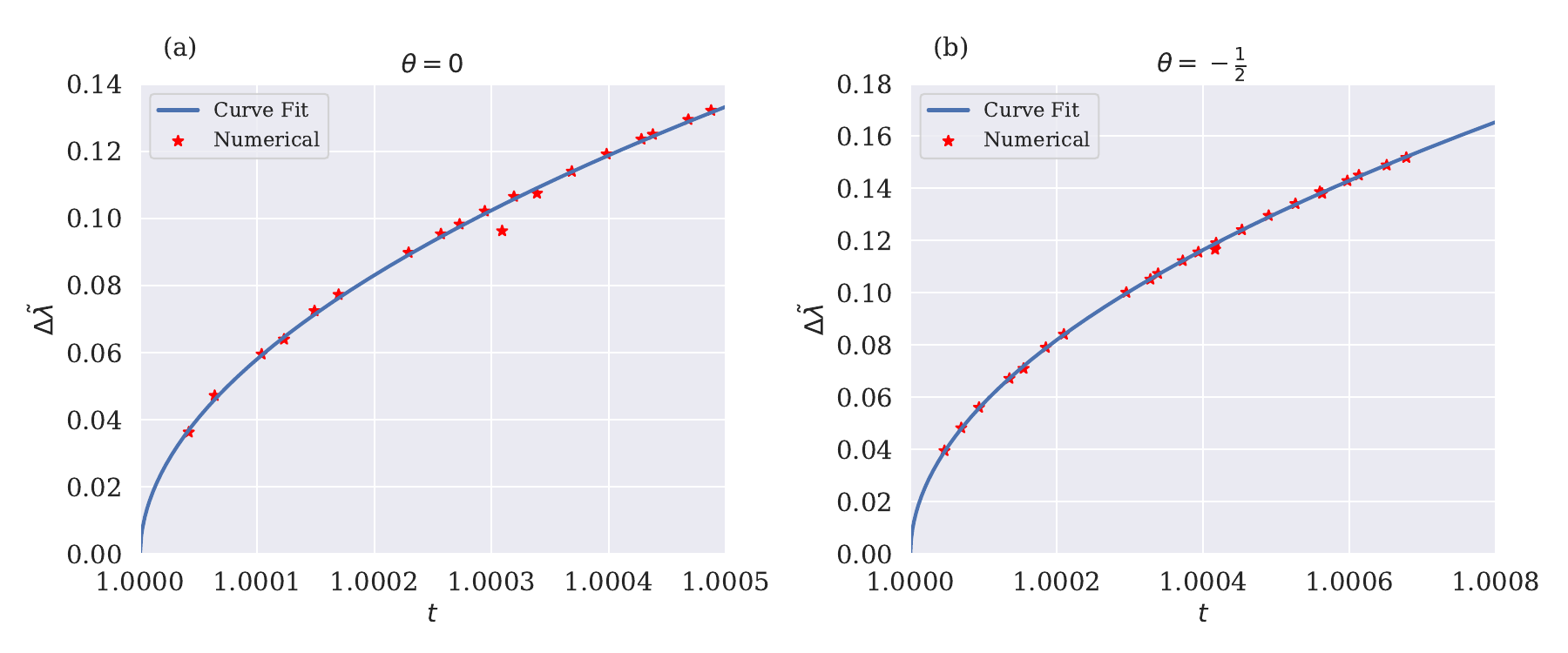}
  \end{center}
  \caption{$ \Delta \tilde \lambda$ vs $\tilde t$ plot for massless particles. Left: Lifshitz case, $\theta = 0$. Right: HSV case, $\theta = -1/2$. The red stars represent data points obtained from numerical calculations, and the blue solid line represents the fitted curve.}%
  \label{curve_fit_photon}%
\end{figure}

In this subsection, we explore the phase transition of HSV black hole using the Lyapunov exponents of null geodesics. Regardless of the geodesic type under consideration, our analysis demonstrates that the Lyapunov exponent remains a reliable probe for phase transitions. Similar to the case of massive particles, our study of massless particles is confined to a specific range of $r_h$ values. The lower limit is set by the extremal black hole condition ensuring the existence of a black hole solution. Conversely, the upper limit is defined by the cessation of photon orbits, a departure from RN-AdS spacetime, where photon orbits are guaranteed for all feasible black hole solutions \cite{Guo:2022kio}. In the HSV spacetime, photon orbits become stable beyond a certain threshold of $r_h$ values. However, our study exclusively focuses on unstable photon orbits.

For a comprehensive understanding of the behavior of Lyapunov exponent, we initially present a 3D plot of $\log_{100} \lambda$ as a function of $\tilde Q$ and $\tilde r_h$ in Figure \ref{3D_Massless} for the Lifshitz case ($\theta =0$). The regions devoid of black hole solutions, determined by $T>0$, have been excluded. The plot reveals that $\lambda$ diverges as $\tilde r_h$ approaches zero. However, due to the absence of unstable photon orbits beyond a certain threshold value of the horizon radius, the Lyapunov exponent $\lambda$ eventually approaches zero. This behavior mirrors that observed in the case of massive particles, distinguishing it from photon orbits in RN-AdS black holes, where $\lambda$ saturates to 1 as $\tilde Q$ or $\tilde r_h$ tends to infinity. In Figure \ref{2d_Massless}, we provide a cross-section of the 3D plot to elucidate the impact of variations in $Q$ and $\theta$ on $\lambda$. In the left panel, the effect of $Q$ is evident for lower $r_h$ values, converging as $r_h$ increases. In the right panel, we illustrate the impact of the HSV parameter for different $\theta$ values. In contrast to the massive case, $\lambda$ decreases as $\theta$ varies from positive to negative. This suggests that the HSV parameter $\theta$ influences massive and massless particles differently.

Inserting $\tilde r_h(\tilde T)$ into Eq. \ref{photon_lambda} yields $\lambda$ as a function of $\tilde T$, as depicted in Figure \ref{fig:lambdaT_massless}. Similar to the massive case, there exists a terminating temperature, at which the unstable null-like circular orbit disappears, and $\lambda$ becomes zero. For $Q< Q_{\text{crit}}$, $\lambda$ is multivalued for the temperature range $\tilde T_1< \tilde T < \tilde T_2$, indicating the coexistence of three black hole solutions. For $Q> Q_{\text{crit}}$, $\lambda$ monotonically decreases and eventually reaches zero. Additionally, the discontinuous change in the Lyapunov exponent $\Delta \tilde \lambda$ is plotted against the temperature $\tilde T$ for massless particles in Figure \ref{curve_fit_photon}, indicating that $\Delta \lambda$ can serve as an order parameter. Near the critical temperature, we find $ \Delta \tilde \lambda = \alpha |t-1|^\delta$, confirming that the critical exponent of $\Delta \lambda$ for the null geodesic case is also $1/2$. Numerically, for the Lifshitz case $\theta=0$, we obtained $\alpha = 6.70366$ and $\delta = 0.515518$, and for the HSV case $\theta=-1/2$, we found $\alpha = 6.1173$ and $\delta = 0.50626$. Thus, it is reasonable to suppose that the critical exponent of $\Delta \tilde \lambda$ is independent of the Lifshitz exponent $z$, the HSV parameter $\theta$, and the geodesic type, provided the black hole undergoes an SBH-LBH phase transition.

\section{Discussion} \label{sec_discussion}

In this article, we investigated the relationship between black hole thermodynamics, specifically phase transitions, and black hole chaos, characterized by the Lyapunov exponent, in the background of Lifshitz and of HSV spacetimes. The Lyapunov exponent plays a crucial role in understanding chaotic systems, providing insights into the divergence and convergence rates of trajectories near the black hole. The study of chaos within the context of black hole physics has garnered significant interest in recent years. On the other hand, an extension of the Gauge/Gravity duality has been witnessed in the construction of gravity models conjectured to be dual to condensed matter systems with anisotropic scaling. Our research aims to interconnect these two aspects of black hole physics, particularly in the realm of black hole thermodynamics. We demonstrate that the Lyapunov exponent effectively characterizes the phase structure of black holes in both hyperscaling and HSV spacetimes, similar to AdS spacetime.

By using the Lyapunov exponent, a measure reflecting the inverse instability timescale associated with geodesic motion,  we have illustrated that the phase configuration within hyperscaling (violating) spacetime is determined by parameters governing circular geodesics. Our primary focus has been on unraveling the details of timelike and null geodesics of particles, aiming to probe the phase structure characterizing black holes within hyperscaling (violating) spacetimes. Given the inherent asymmetry in the scaling of space and time, a careful analysis of geodesics becomes necessary. Therefore, first we have presented a straightforward derivation of the Lyapunov exponent in Section \ref{sec_geodesic}, elucidating its connection to the effective potential of circular geodesics. When the space and time undergo equal scaling, our derivation agrees with the established results from prior studies.

In our investigation, we specifically focused on the thermodynamics of the spherical black holes with $1\leq z \leq 2$ and arbitrary $\theta$ in the canonical ensemble. These black holes exhibit a first-order transition between the small black hole phase and the large black hole phase, similar to the liquid-gas phase transition observed in vdW fluids. Our observations reveal that, when the black hole charge is below the critical value (determined by black hole parameters, including $z$ and $\theta$), the Lyapunov exponents, plotted against Hawking temperature, show three branches, corresponding to three coexisting black hole phases. Conversely, when the charge surpasses the critical threshold, the Lyapunov exponents become single-valued functions of temperature, aligning with one black hole phase. Notably, during the first-order phase transition, the discontinuity in the Lyapunov exponent $(\Delta \lambda)$ serves as an order parameter, effectively characterizing the black hole phase transition. Remarkably, we find that $\Delta \lambda$ exhibits a critical exponent of $1/2$ at the critical point. Given the difficulty of obtaining analytical solutions, we opted for a numerical approach to perform the necessary calculations.

Our findings further validate the proposed connection between Lyapunov exponents and phase transitions in charged black holes extending its applicability to a broader context, beyond AdS spacetime. However, in the selected model, the gauge potentials $A$ and $B$ remain inert in thermodynamics, serving solely to uphold the structure of the asymptotic spacetime and the geometry of the internal space, with their associated charges, such as $Q_F$ and $Q_H$, held fixed. Any modification to these charges would alter the symmetries and geometry of the dual field theory, significantly impacting the holographic interpretation. This observation motivates the exploration of holographic thermodynamics in the realm of HSV spacetime, an area of recent interest in the community \cite{Ahmed:2023snm}. Subsequent analysis could extend to establishing connections between holographic thermodynamics and Lyapunov exponents. An intriguing starting point for this study could involve connecting holographic thermodynamics and Lyapunov exponents, beginning with the AdS spacetime itself.

  \bibliography{BibTex}

\end{document}